%% file: main.tex
\documentclass[useAMS,usenatbib]{mn2e} %MNRAS
\usepackage{graphicx}
\usepackage{amssymb, amsmath}
\usepackage{times}
\usepackage{color}
\usepackage{lscape}
\usepackage{placeins}
\usepackage{stfloats}

\usepackage{lineno}

\usepackage{arydshln}

\usepackage{url}

\def\f{\frac}

\begin{document}

%%%%% Front Matter %%%%%%%%%%%%%%%%%%%%%%%%%%%%%%%%%%%%%%%%%%%%%%%%%%%%%%%%%%%%

\title[LSS $f\sigma_8+CMB$]{Testing deviations from $\Lambda$CDM  with growth rate measurements from 6 Large Scale Structure Surveys at $\mathbf{z=0.06}$ to 1}
\author[Alam et al.] {
    Shadab Alam$^{1,2}$, Shirley Ho$^{1,2}$ and  
    Alessandra Silvestri$^{3}$  \\
    $^{1}$ Departments of Physics, Carnegie Mellon University, 5000 Forbes Ave., Pittsburgh, PA 15217 \\
    $^{2}$ McWilliams Center for Cosmology, Carnegie Mellon University, 5000 Forbes Ave., Pittsburgh, PA 15217 \\
    $^{3}$ Institute Lorentz, Leiden University, PO Box 9506, Leiden 2300 RA, The Netherlands }
    
\date{\today}
\pagerange{\pageref{firstpage}--\pageref{lastpage}}   \pubyear{2015}
\maketitle
\label{firstpage}

%\label{firstpage}
\input{tex/abstract}


%%%%% Main Text %%%%%%%%%%%%%%%%%%%%%%%%%%%%%%%%%%%%%%%%%%%%%%%%%%%%%%%%%%%%%%%
\input{tex/intro}

\input{tex/theory}

\input{tex/data}

\input{tex/systematic}

\input{tex/analysis}

\input{tex/result}

\input{tex/discussion}

%%%%% Bibliography %%%%%%%%%%%%%%%%%%%%%%%%%%%%%%%%%%%%%%%%%%%%%%%%%%%%%%%%%%%%

%\setlength{\bibhang}{2.0em}
%\setlength\labelwidth{0.0em}

\bibliography{Master_Shadab}
\bibliographystyle{mnras}

\label{lastpage}

\end{document}

%% file: tex/abstract.tex
\begin{abstract}
We use measurements  from the Planck satellite mission and  galaxy redshift surveys over the last decade to test three of the basic assumptions of  the standard  model of cosmology, $\Lambda$CDM: the spatial curvature of the universe, the nature of dark energy and the laws of gravity on large scales. We obtain improved constraints on several scenarios that violate one or more of these assumptions. We measure $w_0=-0.94\pm0.17$ (18\% measurement) and $1+w_a=1.16\pm0.36$ (31\% measurement) for models with a time-dependent equation of state, which is an improvement over current best constraints \citep{Aubourg2014}. In the context of modified gravity, we consider  popular scalar tensor models as well as a parametrization of the growth factor. In the case of one-parameter $f(R)$ gravity models with a $\Lambda$CDM background, we constrain $B_0 < 1.36 \times 10^{-5} $ (1$\sigma$ C.L.), which is an improvement by a factor of 4 on the current best \citep{XU2015}. We provide the very first constraint on the coupling parameters of general scalar-tensor theory and stringent constraint on the only free coupling parameter of Chameleon models. We also derive constraints on extended Chameleon models, improving the constraint on the coupling by a factor of 6 on the current best \citep{Hojjati2011} . We also measure $\gamma =  0.612 \pm 0.072$ (11.7\% measurement) for growth index parametrization. We improve all the current constraints by combining results from various galaxy redshift surveys in a coherent way, which includes a careful treatment of scale-dependence introduced by modified gravity.
\end{abstract}

%% file: tex/intro.tex
\section{Introduction}
\label{sec:intro}

Since its development a century ago,  General Relativity (GR) has consistently provided a very successful framework to describe the evolution of our Universe \citep{Peebles1980,Davis83}. Nowadays, the prediction of GR for the  growth of the large scale structure that we observe around us, is reaching great precision as cosmic microwave background (CMB) measurements are providing us with impressively accurate estimates of the cosmological parameters \citep{Planck2015}. Yet, the excitement about the advances of observational cosmology is accompanied by the awareness that we face some major challenges. While the standard cosmological model, based on the laws of GR, provides a very good fit to existing data, it relies on a universe of which we understand only $\sim$5$\%$ of the content. The remaining energy budget comes in the form of dark matter ($\sim$27$\%$), responsible for the clustering of structure,  and the cosmological constant $\Lambda$ \citep{Einstein1915} ($\sim$ 68$\%$), responsible for the phase of accelerated expansion recently entered by the universe. In particular, the physical understanding of cosmic acceleration represents one of the most important  challenges in front of modern physics. While $\Lambda$ is in good agreement with available data, e.g.  baryon acoustic oscillations (BAO) \citep{Eis2005, Cole2005, Hutsi2006, Kazin2010, Percival2010, Reid2010, Eric2014, Anderson2014, Anderson2013}, Supernovae \citep{Perlmutter2003, Conley2011,Goobar2011, Suzuki2012, Rodney2014}, and CMB (\citet{Planck2015}, WMAP9 \citet{Wmap2013}) observations, it suffers from the coincidence and fine tuning problems \citep{Weinberg1989, Carroll2001}.  Several alternatives to $\Lambda$ have been proposed in the two decades since the discovery of cosmic acceleration \citep{Riess1998, Perlmutter1999}, and they can be roughly  divided into two classes. The first class, to which we will refer as \emph{modified gravity} (MG), corresponds to modifications of the laws of gravity on large scales, designed to achieve self accelerating solutions when matter becomes negligible \citep{Silvestri2009, Clifton2012}; alternatively, one can introduce a dynamical degree of freedom, commonly dubbed  \emph{dark energy} (DE; first coined by \cite{Huterer1999}), which is smoothly distributed and starts to dominate the evolution of the Universe at late times \citep{Copeland2006}. 

Undoubtedly, one of the important tasks for modern cosmologists, is to perform precision tests of the standard model of cosmology ($\Lambda$CDM) and identify areas of tension. In a joint effort, one needs also to explore the parameter space of alternative models. Even though with the current constraints from data, any departure from $\Lambda$CDM is likely to be small and challenging to detect,  we are in a unique position to test GR, and the other assumptions of $\Lambda$CDM, to unprecedented precision with modern observational probes. The three basic assumptions of $\Lambda$CDM which are popularly tested are the curvature of the universe, the nature of dark energy and the laws of gravitational interaction on large scales. The curvature of the universe can be explored by  allowing a curvature density parameter, $\Omega_K$, to be different from zero and free to vary. As for the nature of dark energy, we will focus on smoothly distributed models where it suffices to test for the deviation of the equation of state parameter, $w$, from -1, which is the value it assumes if the acceleration is driven by $\Lambda$. We will consider both a constant $w$  as well as a  time-dependent one, resorting to the popular CPL parametrization in terms of $w_0$ and $w_a$, i.e. $w=w_0 + w_a \frac{z}{1+z}$ \citep{Chevallier2001,Linder2003}. Finally, we will explore the nature of gravity  by replacing  GR with various modified gravity models, including Chameleon-type scalar-tensor theories and popular parametrizations of the growth rate. 
All these alternatives that we consider in our analysis, affect, in one way or another,  the rate at which large scale structures grow. Models of smoothly distributed dark energy, which does not cluster, modify only the background dynamics of the universe, but this still has an impact on the rate at which structure forms. On the other hand, models of modified gravity generally modify both the background and perturbation dynamics, leading to a significant effect on the growth rate.

Modern galaxy redshift surveys, have successfully measured the growth rate using Redshift Space Distortions (hereafter RSD ; \citet{Kaiser87}), which is the distortion induced in the galaxy correlation function by the peculiar velocity component of the galaxy redshift. Hence, on linear scales, RSD offers a handle both on  the distribution of matter over-density and peculiar velocity of galaxies. Recent galaxy redshift surveys have provided the measurement of $f\sigma_8(z)$ up to redshift  $z=0.8$ , where $f$ is the growth rate, i.e. the logarithmic derivative of the growth factor, and $\sigma_8$ is the rms amplitude of matter fluctuations in a sphere of radius 8 h$^{-1}$Mpc. In this paper, we will test all the three assumptions of $\Lambda$CDM listed above using the Planck CMB measurement \citep{Planck2013} and latest RSD measurement  from BOSS CMASS \citep{Alam2015},  SDSS LRG \citep{SDSSLRG2012}, 6dFGRS\citep{6dFGRS}, 2dFGRS  \citep{2dFGRS}, WiggleZ \citep{Blake2011} and VIMOS Public Extragalactic Redshift Survey (VIPERS,\cite{Vipers}) . It is difficult to use the measurement from different surveys as they have different assumptions. We have looked into these assumptions and possible systematic while combining results from the different survey and also proposed a way to test scale dependence for modified gravity models using these results.

%% file: tex/theory.tex
\section{Theory}
In exploring the power of RSD data to constrain deviations from the standard cosmological scenario,  we consider several alternative models,  divided into dark energy models that modify the background expansion history without introducing any clustering degree of freedom, and those that instead modify only the dynamics of perturbations while keeping the background fixed to $\Lambda$CDM.  In the former case we consider one and two parameter extensions of the standard scenario, corresponding to different equations of state for dark energy or a non zero spatial curvature. More specifically we consider: a $w$CDM universe, where the equation of state for dark energy is a constant parameter that can differ from the $\Lambda$CDM value  $w=-1$; a $(w_0,w_a)$CDM universe, in which the equation of state for dark energy is a function of time and is approximation to exact solutions of the scalar field equation of motion, i.e.  the Chevallier-Polarski-Linder (CPL) parameterization $w=w_0+w_a(1-a)$; a $o\Lambda$CDM universe which can have a spatial curvature different from zero, parameterized in terms of the corresponding fractional energy density $\Omega_{K}$. In the case of models that modify the equations for the evolution of perturbations, we analyze Chameleon-type scalar-tensor theories, $f(R)$ gravity and a time dependent parametrization of the growth rate. 

We use the publicly available Einstein-Boltzmann solver MGCAMB \citep{Hojjati2011}\footnote{ \url{http://www.sfu.ca/~aha25/MGCAMB.html}} to evolve the dynamics of scalar perturbations and obtain predictions to fit to our data set for all the models considered, except for the $(w_0,w_a)$CDM case. This latter needs to be treated instead through the PPF module \citep{Fang2008} in CAMB \footnote{\url{http://camb.info}}. While the implementation of the non-clustering dark energy models is trivial, in the following we shall describe in more detail the implementation of the modified gravity models.

\subsection{Scalar-tensor theories} 
Going beyond simple extensions of the standard model and non-clustering dark energy models, one needs to take into consideration also the modifications to the equations for cosmological perturbations. Given the  cosmological probes that we consider in our analysis, it suffices for us to focus on linear scalar perturbations. In this context, it is possible to generally parametrize deviations from the standard cosmological scenario in the dynamics of perturbations by mean of two functions of time and scale introduced in the set of Einstein and Boltzmann equations for metric and matter perturbations. More precisely, one can write the Poisson and anisotropy equations as follows:
 \begin{equation}\label{mu}
 k^2\Psi=-\frac{a^2}{2M_P^2}\mu(a,k)\rho\Delta \ ,\,\,\,\,\frac{\Phi}{\Psi}=\gamma_{slip}(a,k) \,,
 \end{equation}
where $\rho\Delta\equiv\rho\delta+3\frac{aH}{k}(\rho+P)v$  is the comoving density perturbation of matter fields and we have selected the conformal Newtonian gauge with $\Psi$ and $\Phi$ representing the perturbation to respectively the time-time and space-space diagonal component of the metric. And then combine them with the unmodified Boltzmann equations for matter fields. 

We shall focus on scalar-tensor theories where the metric and the additional scalar degree of freedom  obey second order equations of motion and will adopt the parametrization introduced in~\citep{Bertschinger2008} (BZ) to describe the corresponding form of $(\mu,\gamma_{slip})$, i.e.:
\begin{eqnarray}\label{mu_gamma_BZ}
\mu =\f{1+\beta_1\lambda_1^2\,k^2a^s}{1+\lambda_1^2\,k^2a^s}\,,\nonumber\\
\gamma_{\rm slip}=\f{1+\beta_2\lambda_2^2\,k^2a^s}{1+\lambda_2^2\,k^2a^s}
\end{eqnarray}
where we have adopted the convention of~\citep{Zhao2009} and $\beta_1,\beta_2$ are dimensionless contants representing couplings, $\lambda_1,\lambda_2$ have dimensions of length and $s>0$ to ensure that at early times GR is recovered. This parametrization gives a very good representation of scalar-tensor theories in the quasi-static regime, where time derivates of the perturbations to the metric and scalar degree of freedom are neglected with respect to their spatial gradients on sub-horizon scales~\citep{Zhao2009,Hojjati2012,DeFelice2010,Amendola2013,Silvestri2013}. This is a good approximation given the observables that we are considering. Additionally, ~(\ref{mu_gamma_BZ}) sets the evolution of the characteristic lenghtscales of the models to a power law in the scale factor. This is of course a choice of parametrization for the time dependence of the mass scale of the scalar degree of freedom, and other choices are possible. Nevertheless, as we will discuss in the following, it is a good approximation for several scalar-tensor models, and data  are not that sensitive to the specific choice of the time dependence. 

Equations~(\ref{mu_gamma_BZ}) are built-in in MGCAMB and allow to easily extract predictions for scalar-tensor models on a $\Lambda$CDM background for different observables, including the growth rate.

\subsection{Chameleon models}
Chameleon models are a class of scalar-tensor theories for which the additional scalar field has a standard kinetic term and is conformally coupled to matter fields as follows:
\begin{eqnarray}\label{Einstein_action_text}
S=&&\int d^4x\sqrt{-\tilde{g}}\left[\f{M_P^2}{2}\tilde{R}-\f{1}{2}\tilde{g^{\mu\nu}}(\tilde{\nabla}_{\mu}\phi)\tilde{\nabla}_{\nu}\phi-V(\phi)\right]\nonumber\\
&&+S_i\left(\chi_i,e^{-\kappa\alpha_i(\phi)}\tilde{g}_{\mu\nu}\right)\,,
\end{eqnarray}
where $\alpha_i(\phi)$ is the coupling between the scalar field $\phi$ and the i-th matter species. The coupling(s) in general can be a non-linear function(s) of the field $\phi$; however, since the value of the field $\phi$ typically does not change significantly on the time scales associated to the epoch of structure formation, we will assume it to be linear in $\phi$. Since we are dealing with clustering of matter in the late universe, it is safe to consider one coupling, i.e. to dark matter; that amounts to neglecting differences between baryons and dark matter, or simply neglecting baryons, which is safe for the observables under consideration.

 In the quasi-static regime, $(\mu,\gamma_{slip})$ for Chameleon-type theories can be well represented  by a simplified version of~(\ref{mu_gamma_BZ})  for which:

\begin{equation}\label{Ch_cond}
1+\frac{1}{2}\left(\frac{d \alpha}{d\phi}\right)^2=\beta_1=\frac{\lambda_2^2}{\lambda_1^2}=2-\beta_2\frac{\lambda_2^2}{\lambda_1^2}\,,\,\,\, 1\leq s\leq 4
\end{equation}

Therefore the effects of Chameleon-type theories on the dynamics of linear scalar perturbations on sub-horizon regimes can be described with good accuracy in terms of  three parameters: $\{\beta_1,\lambda_1,s\}$. The last condition in~(\ref{Ch_cond}) is broadly valid for models with runaway and tracking type potentials~\citep{Zhao2009}. Following a convention which is commonly used for $f(R)$ theories, let us express the lengthscale $\lambda_1^2$ in terms of a new parameter $B_0$, which corresponds to the value of the inverse mass scale today in units of the horizon scale~\citep{Song2007}:
\begin{equation}\label{B0}
B_0\equiv\frac{2H_0^2\lambda_1^2}{c^2},
\end{equation}
so that we will work with $\{\beta_1,B_0,s\}$.

Let us notice that Chameleon theories as defined in action (\ref{Einstein_action_text}), have necessarily $\beta_1\geq1$. However, in previous analysis of Chameleon models under the BZ parametrization, such theoretical prior has not been generally  imposed and a wider range of $\beta_1$ has been explored (see e.g.~\citep{Hojjati2011,DiValentino2012}). Hence, in our analysis we will consider both the case with $\beta_1>1$ and the case for which $\beta_1$ is allowed to be smaller than unity, to facilitate comparison. We will refer to the former as the \emph{Chameleon}  model, and the latter as the extended Chameleon model (\emph{eChameleon}). We shall emphasize that we consider the eChameleon as a purely phenomenological  model within (\ref{mu_gamma_BZ}), without linking it to action (\ref{Einstein_action_text}), since it would not be viable case of the latter. While the eChameleon might correspond to a very special subcase of the parametrization~(\ref{mu_gamma_BZ}), it still represents a possible choice for $(\mu,\gamma_{slip})$ and, as we will discuss in Section~\ref{sec:results}, it will be interesting to see what data can say about it.

\subsection{$f(R)$ gravity}
$f(R)$ theories of gravity correspond to the simple modification of the Einstein-Hilbert action by the addition of a nonlinear function of the Ricci scalar. In the past decade they have been extensively explored as candidate models for cosmic acceleration (see e.g.~\citep{Silvestri2009,DeFelice2010} and references therein). They represent a subcase of the larger class of models described by action~(\ref{Einstein_action_text}), corresponding to a universal fixed coupling $\alpha_i=\sqrt{2/3}\,\phi$ and are therefore well represented in the quasi-static regime by the functions~(\ref{mu_gamma_BZ}) and conditions~(\ref{Ch_cond}). However, the fixed coupling $\alpha_i=\sqrt{2/3}\,\phi$ implies that $\beta_1=4/3$ and viable $f(R)$ models that closely mimic $\Lambda$CDM have been shown to correspond to $s\sim 4$~\citep{Zhao2009,Hojjati2012}. Therefore the number of free parameters in Eqs.~(\ref{mu_gamma_BZ}) can be effectively reduced to $\lambda_1$, which is then expressed in terms of $B_0$. The latter is in fact the only free parameter needed to label the family of $f(R)$ models that reproduce a given expansion history, in our case the $\Lambda$CDM one, and can be usually reconstructed via the so-called designer approach~\citep{Song2007,Pogosian2008}. 
Alternatively, one could adopt the recently developed EFTCAMB package for an exact implementation of designer $f(R)$ models that does not rely on the quasi-static approximation \citep{Hu2014,Raveri2014} \footnote{\url{http://wwwhome.lorentz.leidenuniv.nl/~hu/codes/}}. The latter method allows to choose different bacgkround histories, however for the data and cosmology involved in our analysis, MGCAMB provides enough accuracy.

\subsection{Growth index parametrization of the growth rate}
In the cosmological concordance model, as well as in non-clustering dark energy models, the growth rate of structure is well approximated by:
\begin{equation}\label{growth_rate}
f\equiv\frac{d\ln\delta_m}{d\ln{a}}\approx\Omega_m(a)^{6/11}
\end{equation}
where $\Omega_m(a)\equiv\rho_m(a)/3M_P^2H^2(a)$, $\rho_m$ is the background density of matter and $\delta_m\equiv\delta\rho_m/\rho_m$.
This inspired the following parametrization for deviations in the growth of structure~\citep{Wang1998,Linder2005,Linder2007}
 \begin{equation}\label{Linder}
 f=\Omega_m(a)^{\gamma}
 \end{equation}
 where $\gamma$ is commonly referred to as growth index (not to be confused with the $\gamma_{slip}$ defined above, which represents instead the gravitational slip). 

The idea behind this parametrization is that of capturing independently in $\Omega_m$ and $\gamma$ the information from, respectively, the expansion and the growth history.  Since in our analysis we fix the background to $\Lambda$CDM, $\Omega_m(a)$ is determined by that and the only parameter of interest will be $\gamma$. While for models of modified gravity and clustering dark energy in general $\gamma$ will be a function of time and scale, in several cases for the regime of interest it can still be safely approximated by a constant, which can differ significantly from the $\Lambda$CDM value. See~\citep{Linder2007} for more details and some forms of $\gamma$ in alternative theories of gravity. 

In our analysis we will assume $\gamma$ is constant and explore constraints on it after extracting predictions for the CMB and growth of structure from MGCAMB.

%% file: tex/data.tex
\section{Observations} 
\label{sec:data}

\begin{figure}
\includegraphics[width=0.5\textwidth]{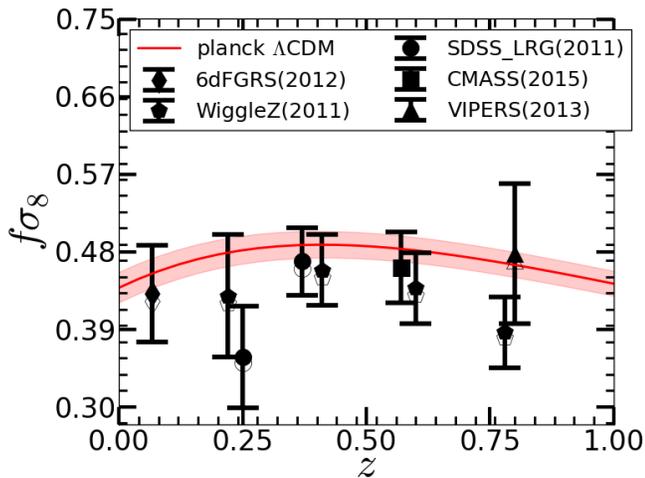}
\caption{The measured $f\sigma_8$ from different surveys covering redshift range $0.06<z<0.8$. The empty markers represent the reported measurement of $f\sigma_8$ and the filled markers are for the corrected values for Planck Comsology. The red band shows the Planck $\Lambda$CDM $1\sigma$ prediction. }
\label{fig:fs8z_data}
\end{figure}

We use measurements of CMB angular power spectrum ( $C_l$ ) from Planck 2013 \citep{Planck2014..15P} combined with the measurement of $f(z)\sigma_8(z)$ from various redshift surveys covering between $z=0.06$ to $z=0.8$ listed in Table \ref{tbl:fs8} as our main data points. The Figure \ref{fig:fs8z_data} shows the measurements used with and without corrections and Planck 2013 prediction. We briefly describe each of the surveys and $f\sigma_8$ measurements in the following sections.

\begin{table}
\begin{center}
\caption{Measurement of $f(z)\sigma_8(z)$ from various galaxy redshift surveys covering redshift between 0.06 to 0.8.}
\label{tbl:fs8}
\begin{tabular}{llll}
\hline
z              & $f\sigma_8(z)$ & 1/k[h /Mpc] & Survey  \\
\hline
\hline
0.067   & $0.42 \pm 0.05$     & 16.0 --30   & 6dFGRS(2012) \\
0.17     & $0.51 \pm 0.06$     &  6.7  -- 50  & 2dFGRS(2004)\\
0.22     & $0.42 \pm 0.07$     &  3.3  -- 50  & WiggleZ(2011) \\
0.25     & $0.35 \pm 0.06$     &  30 -- 200  & SDSS LRG (2011) \\
0.37     & $0.46 \pm 0.04$     &  30 -- 200  & SDSS LRG(2011)\\
0.41     & $0.45 \pm 0.04$     &  3.3 -- 50   & WiggleZ(2011)\\
0.57     & $0.462 \pm 0.041$     &  25 --130   & BOSS CMASS\\
0.6       & $0.43 \pm 0.04$     &  3.3 -- 50   & WiggleZ(2011) \\
0.78     & $0.38 \pm 0.04$     &  3.3 -- 50   & WiggleZ(2011)\\
0.8       & $0.47 \pm 0.08$     &  6.0 -- 35   & Vipers(2013) \\
\end{tabular}
\end{center}
\end{table}

\subsection{6dFGRS}
The 6dFGRS (6 degree Field Galaxy Redshift Survey has observed 125000 galaxies in near infrared band across 4/5th of southern sky \cite{Jones2009}. The surveys covers redshift range $0<z<0.18$, and has an effective volume equivalent to 2dFGRS \citep{2dFGRS} galaxy survey. The RSD measurement was obtained using a subsample of the survey consisting of 81971 galaxies \citep{6dFGRS}. The measurement of $f\sigma_8$ was obtained by fitting 2D correlation function using streaming model and fitting range 16-30 Mpc/h. The Alcock-Paczynski effect \cite{AP79} has been taken into account and it has a negligible effect \citep{6dFGRS}. The final measurement uses WMAP7 \citep{Wmap2013} likelihood in the analysis. To be able to use this $f\sigma_8$ measurement we need to account for the transformation to the Planck best fit cosmology \citep{Planck2013}. 

\subsection{2dFGRS}
The 2dFGRS (2 degree Field Galaxy Redshift Survey) obtained spectra for 221414 galaxies in visible band on the southern sky \citep{Colless2003}. The survey covers redshift range $0<z<0.25$ and has a effective area of 1500 square degree. The RSD measurement was obtained by linearly modeling the observed distortion after splitting the over-density into radial and angular components \citep{2dFGRS}. The parameters were fixed at different values $n_s=1.0$ ,$H_0=72$. The results were marginalized over power spectrum amplitude and $b\sigma_8$.  We are not using this measurement in our analysis for two reasons. First, the survey has a huge overlap with 6dFGRS which will lead to a strong correlation between the two measurements. Second, the cosmology assumed is quite far from WMAP7 and Planck which may cause our linear theory approximation used to shift the cosmology might fail.

\subsection{WiggleZ}
The WiggleZ Dark Energy Survey is a large scale galaxy redshift survey of bright emission line galaxies. It has obtained spectra for nearly 200,000 galaxies. The survey covers redshift range $0.2<z<1.0$, covering effective area of 800 square degrees of equatorial sky \citep{WiggleZ}. The RSD measurement was obtained using a sub-sample of the survey consisting of 152,117 galaxies. The final result was obtained by fitting the power spectrum using \citet{Jennings2011} model in four non-overlapping slices of redshift. The measured growth rate is $f\sigma_8(z)=( 0.42 \pm 0.07, 0.45 \pm 0.04, 0.43 \pm 0.04, 0.38 \pm 0.04)$ at effective redshift $z=(0.22, 0.41, 0.6, 0.78)$ with non-overlaping redshift slices of $z_{slice}=([0.1,0.3],[0.3,0.5],[0.5,0.7],[0.7,0.9])$ respectively. We can assume the covariance between the different measurements to be zero because they have no volume overlap.

\subsection{SDSS-LRG}
The Sloan Digital Sky Survey (SDSS) data release 7 (DR7) is a large-scale galaxy redshift survey of Luminous Red Galaxies (LRG) \citep{Eisenstein2011}. The DR7 has obtained spectra of 106,341 LRGs, covering 10,000 square degree in redshift range $0.16<z<0.44$. The RSD measurement was obtained by modeling monopole and quadruple moment of galaxy auto-correlation function using linear theory. The data was divided in two redshift bins: $0.16<z<0.32$ and $0.32<z<0.44$. The measurements of growth rate are $f\sigma_8(z)=(0.3512 \pm 0.0583, 0.4602 \pm 0.0378)$ at effective redshift of 0.25 and 0.37 respectively \citep{SDSSLRG2012}. These measurements are independent because there is no overlapping volume between the two redshift slices.

\subsection{BOSS CMASS}
Sloan Digital Sky Survey (SDSS) Baryon Oscillation Spectroscopic Survey (BOSS; \citet{Dawson2013}) targets high redshift ($0.4 < z < 0.7$) galaxies using a set of color-magnitude cuts. The growth rate measurement uses the CMASS (Reid et. al. in prep, \citet{Anderson2014}) sample of galaxies from Data Release 11 \citep{Alam2014}. The CMASS sample has 690,826 Luminous Red Galaxies (LRGs) covering 8498 square degrees in the redshift range $0.43<z<0.70$, which correspond to an effective volume of 6 Gpc$^{3}$. The $f\sigma_8$ is measured by modeling the monopole and quadruple moment of galaxy auto-correlation using Convolution Lagrangian Perturbation Theory (CLPT; \citet{Carlson12}) in combination with Gaussian Streaming model \citep{Wang13}. The reported measurement of growth rate is $f\sigma_8=0.462 \pm 0.041$ at effective redshift of $0.57$ \citep{Alam2015}. 

We are also using the combined measurement of growth rate ($f\sigma_8$), angular diameter distance ($D_A$) and Hubble constant ($H$) measured from the galaxy auto correlation in CMASS sample at an effective redshift of 0.57 \citep{Alam2015}. The measurement and its covariance are given below and it's called eCMASS.

\begin{equation*}
f\sigma_8 = 0.46 ,
D_A = 1401 ,
H = 89.15
\end{equation*}

\begin{equation}
\setlength{\arraycolsep}{0.5pt}
C_{eCMASS} =
\scriptscriptstyle
\left(
\begin{matrix}
0.0018 & -0.6752 & -0.1261\\
-0.6752 & 550.61 & 45.881 \\
-0.1261 & 45.881 & 14.019
\end{matrix}
\right)
\label{eq:eCMASS}
\end{equation}

\subsection{VIPERS}
VIMOS Public Extragalactic Redshift Survey (VIPERS,\cite{Vipers}) is a high redshift small area galaxy redshift survey.  It has obtained spectra for 55,358 galaxies covering 24 square degree in the sky from redshift range $0.4<z<1.2$.  The measurement of growth factor uses 45,871 galaxies covering the redshift range $0.7<z,1.2$.
The $f\sigma_8$  measurement is obtained by modeling the monopole and quadruple moments of galaxy auto-correlation function between the scale 6 h$^{-1}$Mpc and 35 h$^{-1}$Mpc. They have reported $f\sigma_8=0.47 \pm 0.08$ at effective redshift of 0.8.  The perturbation theory used in the analysis has been tested against $N$-body simulation and shown to work at mildly non-linear scale below 10 h$^{-1}$Mpc \citep{Vipers}.

\subsection{Planck {\it CMB}}
Planck is a space mission dedicated to the measurement of CMB anisotropies.  It is the third-generation of all sky CMB experiment following COBE and WMAP. The primary aim of the mission is to measure the temperature and polarization anisotropies over the entire sky. The Planck mission provides a high resolution map of CMB anisotropy which is used to measure the cosmic variance limited angular power spectrum $C^{TT}_\ell$ at the last scattering surface. The Planck measurements helps us constrain the background cosmology to unprecedented precision \citep{Planck2013,PlanckXVI, Planck2015}. We are using the CMB measurements from Planck satellite in order to constrain cosmology. We have assumed that Planck measurements is independent of the measurement of growth rate from various galaxy redshift surveys.

\subsection{Correlation Matrix}
\label{sec:corr}
\begin{figure}
\includegraphics[width=0.5\textwidth]{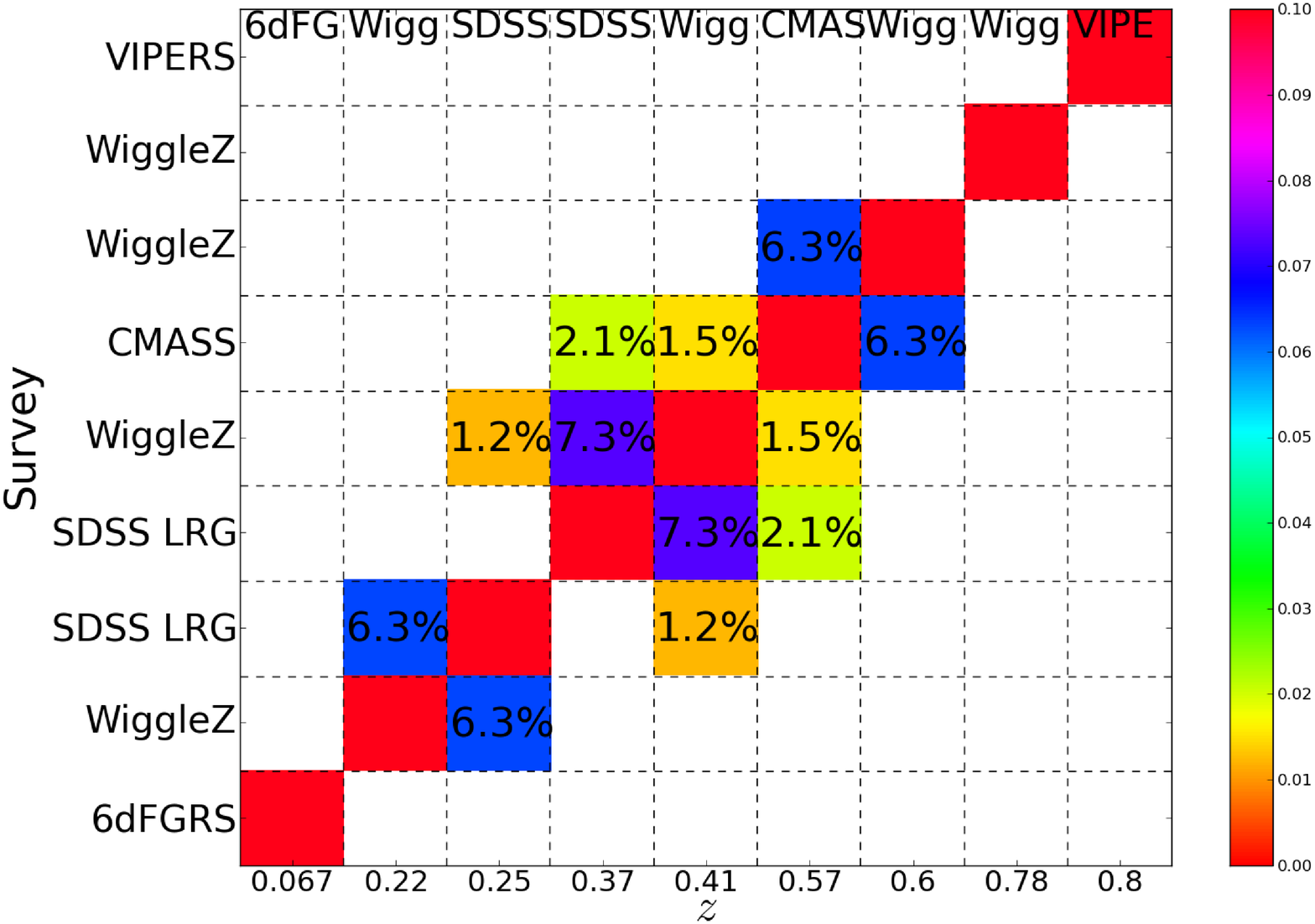}
\caption{Correlation matrix between all the measurements used in our analysis. We have estimated the correlation as the fraction of overlap volume between two survey to the total volume of the two surveys combined.}
\label{fig:corrmat}
\end{figure}

We use the measurements of $f\sigma_8$ from 6 different surveys. Although these surveys are largely independent, and in some cases they probe different biased tracers, they are measuring inherently the same matter density field. Therefore, the parts of the survey observing same volume of sky cannot be treated as independent. We have predicted an upper limit to the overlap volume using the data from different surveys. We have estimated the fractional overlap volume between any two samples as the ratio of the overlap volume to the total volume of the two samples. We estimate the correlation between two measurements as the fractional overlap volume between the two measurements. Figure \ref{fig:corrmat} shows our estimate of the correlation between the surveys. The four measurements of WiggleZ survey cover the redshift range between 0.1 to 0.9 and hence show most correlation with other measurements like SDSS LRG and CMASS in the same redshift range. 

%% file: tex/systematic.tex
%%%%%%%%%%%%%%%%%%%%%%%%%%%%%%%%%%%%%%%%%%%%%%%%%%%%%%%%%%%%%%%%%%%%%%%%%%%%%%%%

\section{Potential Systematics} 
The collection of $f\sigma_8$ data points that we are using in this analysis contain measurements from several different surveys, obtained during the last decade, each with a different pipeline. Furthermore, often the latter implicitly assumes a GR modeling, which does not take into account the different predictions for the growth factor in modified theories of gravity. It is important to account for some crucial differences in order to use these measurements in our analysis. We have looked at following different aspects of measurements and theoretical prediction before using them in our analysis.

\subsection{Fiducial Cosmology of the growth rate ($f\sigma_8$)}
\label{sec:fidcosmo}
The measurements of $f\sigma_8$ have been obtained over the time when we had transition from WMAP best fit cosmology \citep{Hinshaw2013} to the Planck  best fit cosmology \citep{Planck2013}. Since we are using Planck likelihood \citep{Planck2014..15P} in our analysis, we have decided to convert all the measurements to Planck cosmology.
The 3 dimensional correlation function can be transformed from WMAP to the Planck cosmology using Alcock-Paczynski effect (\citet{AP79}),
\begin{equation}
\xi_{planck}(r_\parallel ,r_\perp,\phi) =\xi_{WMAP}(\alpha_\parallel r_\parallel ,\alpha_\perp r_\perp,\phi),
\end{equation}

where $\alpha_\parallel$ is the ratio of the hubble parameters ($\alpha_\parallel=H_{WMAP}/H_{planck} $) and $\alpha_\perp$ is the ratio of the angular diameter distances ( $\alpha_\perp=D_A^{planck}/D_A^{WMAP}$). The $r_\parallel$, $r_\perp$ are pair separations along the line of sight and perpendicular to the line of sight and $\phi$ is the angular position of pair separation vector in the plane perpendicular to the line of sight from a reference direction. In practice the correlation function is isotropic along $\phi$.
We can calculate the corresponding power spectrum by applying Fourier transform to correlation function.

\begin{align}
P_{planck}(k_\parallel,k_\perp,k_\phi) &= \int dr_\parallel dr_\perp r_\perp d\phi \xi_{planck}(r_\parallel ,r_\perp,\phi) e^{-i\vec{k}.\vec{r}} \\
     &= \int dr'_\parallel dr'_\perp \frac{r'_\perp}{\alpha_\parallel \alpha_\perp^2} \xi_{WMAP}(r'_\parallel ,r'_\perp,\phi) e^{-i\vec{k'}.\vec{r'}} \\
     &= \frac{P_{WMAP}(k_\parallel/\alpha_\parallel ,k_\perp/ \alpha_\perp,k_\phi)}{\alpha_\parallel \alpha_\perp^2}
\end{align}

The Kaiser formula for RSD gives the redshift space correlation function as $P_g^s(k,\mu) = b^2 P_m(k) (1 + \beta \mu^2)^2$ \citep{Kaiser87}. Using the linear theory kaiser prediction and the above approximation between WMAP and planck power spectrum, we can get a relation to transform the growth function from WMAP to planck cosmology.

\begin{align}
\frac{1+\beta_{planck} \mu'^2}{1+\beta_{WMAP} \mu^2} &= C \sqrt{\frac{P_{planck}(k',\mu')}{P_{WMAP}(k,\mu)}} \\
          &= C \sqrt{\frac{1}{\alpha_\parallel \alpha_\perp^2}} 
\label{eqn:beta1}
\end{align}

where $C$ is the ratio of isotropic matter power spectrum with WMAP and planck cosmology  integrated over scale used in $\beta$ measurement.

\begin{equation}
C = \int_{k_1}^{k_2} dk \sqrt{\frac{P_{WMAP}^{m}(k)}{P_{planck}^{m}(k')}} 
\end{equation}

When right hand side of equaion(\ref{eqn:beta1}) is close to 1, then we can approximate the above equation as follows:

\begin{equation}
\beta_{planck} = \beta_{WMAP} C \frac{\mu^2}{\mu'^2} \sqrt{\frac{1}{\alpha_\parallel \alpha_\perp^2}} 
\label{eqn:beta2}
\end{equation}

The ratio $\frac{\mu^2}{\mu'^2}$ can be obtained using simple trigonometry which gives following equations, where the last equation is aprroximation for $\alpha_\parallel^2 \approx \alpha_\perp^2$.

\begin{equation}
\frac{\mu^2}{\mu'^2} =\frac{1}{\alpha_\perp^2} \left( \alpha_\parallel^2 +(\alpha_\perp^2 -\alpha_\parallel^2) \mu^2 \right) \approx \left(\frac{\alpha_\parallel}{\alpha_\perp}\right)^2
\label{eqn:muratio}
\end{equation}

We can substitute equation (\ref{eqn:muratio}) in equation (\ref{eqn:beta2}) in order to get the required scaling for $f$(growth factor) assuming that bias measured is proportional to the $\sigma_8$ of the cosmology used.

\begin{equation}
\beta_{planck}=\beta_{WMAP} C \left(\frac{\alpha_\parallel}{\alpha_\perp^2} \right)^{(3/2)}
\end{equation}

\begin{equation}
{f\sigma_8}_{planck} = {f\sigma_8}_{WMAP} C \left(\frac{\alpha_\parallel}{\alpha_\perp^2} \right)^{(3/2)} \left(\frac{\sigma_8^{planck}}{\sigma_8^{WMAP}} \right)^2
\label{eqn:fs8}
\end{equation}

We have tested prediction of equation (\ref{eqn:fs8}) against the measurement of $f\sigma_8$ reported in Table 2 of \citet{Alam2015} at redshift 0.57 using both Planck and WMAP cosmology.

\subsection{Scale dependence}\label{scale_dep}
\label{sec:scaledep}

General Relativity predicts a scale independent growth factor. One of the important features of the modified gravity theories we are considering is that they predict a scale dependent growth factor which has a transition from high to low growth at certain scale which depends on the redshift $z$  and the model parameters. 
The mesurements we use from the different surveys, assumes a scale-independent $f\sigma_8$ and uses characteristic length scale while analyzing data. In order to account for all these effects we have done our analysis in two different ways. In the first method, we assume that the measurements corresponds to an effective k and in the second method, we treat the average theoretical prediction over range of $k$ used in  $f\sigma_8$  analysis.

Figures \ref{fig:Chambeta} and \ref{fig:fR} show the parameters constraint for Chameleon models and $f(R)$ gravity. The grey and red contours result from using two different model predictions to test the scale dependence. The grey contours correspond to the model where we  average $f\sigma_8$ over $k$ used in respective $f\sigma_8$ analysis and red contours correspond to $f\sigma_8$ evaluated at $k=0.2h$ Mpc$^{-1}$. It is evident from the plots that, at the current level of uncertainty, we obtain very similar constraint and hence do not detect any significant effect of scale dependence of $f\sigma_8$.

\subsection{Other systematics}

The measurements of $f\sigma_8$ are reported at the mean redshift of the surveys. But the galaxies used have a redshift distribution which in principle can be taken into account by integrating the theoretical prediction. This should be a very small effect because the $f\sigma_8$ is relatively smooth and flat (see Figure \ref{fig:fs8z} and \citet{Huterer2013arXiv1309.5385H})for the redshift range of the survey and also because the survey window for every individual measurement is small. Another important point is the assumption of GR based modeling for the measurement. We have looked at the modeling assumption for each of the measurements. All measurements of $f\sigma_8$ except WiggleZ and VIPERS, allow the deviation from GR through Alcock-Paczynski effect \citep{AP79} which  justifies our use of {\it modified gravity} models. The inclusion of AP in WiggleZ and VIPERS will marginally increase the error on the measurements. Different surveys use different ranges of scale in the RSD analysis. This will be important especially while analyzing {\it modified gravity} models.  To account for the different scales used we evaluate the prediction for each survey averaged over the scale used in the respective analysis. 

%
%\subsection{redshift of the mesurement}
%The different survey has different redshift coverage and the mesurement are reported at effective redshift \textcolor{red}{ maybe it will be good to have a plot of redshift coverage for different survey}. The $f\sigma_8$ has similar redshift evolution for all the modified gravity theory. We have are using effective redshift of the mesurement to evaluate our theoretical prediction. We have also looked at the effect of averaging the theoretical prediction over the redshift coverage of the survey with the weighting as galaxy number density o the survey. \textcolor{red}{Either explain the result of this here or refer to the section where this will be explained}

%\subsection{Assumption of GR in the survey analysis}
%Generally all these measurement are independent of model of gravityand should be good to test alternate gravity model \textcolor{red}{explain how?}

%% file: tex/analysis.tex
%%%%%%%%%%%%%%%%%%%%%%%%%%%%%%%%%%

\section{Analysis} 
\label{sec:analysis} 

We have measurement of $f\sigma_8$ from various surveys covering redshift range 0.06-0.8 (see Table: \ref{tbl:fs8}). We first correct these measurements for the shift from WMAP cosmology to planck cosmology as described in section(\ref{sec:fidcosmo}). The next step is to evaluate prediction from different modified gravity theories by evolving a full set of linear perturbation equations. The theoretical predictions for $f\sigma_8$ is generally scale and redshift dependent (see section \ref{sec:scaledep}). Therefore, we consider two cases for theoretical prediction: 1) evaluate $f\sigma_8$ at effective k and 2) evaluate $f\sigma_8$ averaged over range of $k$ used in measurements. We also predict $C_l^{TT}$ for different modified gravity theories. Finally we define our likelihood, which consists of three parts one  by matching planck temperature fluctuation $C_l^{TT}$ , second by matching growth factor from Table \ref{tbl:fs8} and third by using eCMASS data as shown in equation \ref{eq:eCMASS}. Therefore, we define the likelihood as follows:

\begin{align}
& \mathcal{L} = \mathcal{L}_{planck}  \mathcal{L}_{f\sigma_8} \mathcal{L}_{eCMASS} \\
& \mathcal{L}_{f\sigma_8} = e^ {-\chi^2_{f\sigma_8}/2} \\
& \chi^2_{f\sigma_8} = \Delta f \sigma_8 C^{-1} \Delta f\sigma_8^{T}
\end{align}

The $\Delta f\sigma_8$ is the deviation of the theoretical prediction from the mesurement and $C^{-1}$ is the inverse of covariance which has diagonal error for different surveys and correlation between measurement as described in section[\ref{sec:corr}].  Note that we do not include $f\sigma_8$ from CMASS while using eCMASS with $f\sigma_8(z)$ to avoid double counting. This likelihood is sampled using modified version of COSMOMC \citep{cosmomc,Hojjati2011}. We sample over 6 cosmology parameters \{$\Omega_b h^2, \Omega_c h^2, 100\Theta_{MC} ,\tau, n_s, log(10^{10} A_s) $\} and all 18 planck nuisance parameters as described in \citet{Planck2013} with the respective extension parameters or modified gravity parameters. 
The priors we have used on all the parameters are the same as the priors in \citet{Planck2013}  and the priors we used on the parameters of modified gravity model are given in Table [\ref{tbl:LCDM-ext}].

%% file: tex/result.tex
\section{Results} 
\label{sec:results}

\begin{table}
\begin{center}
\caption{The list of extension parameters for all the models used in our analysis. For each parameter we provide their symbol, prior range, central value and $1\sigma$ error.}
\label{tbl:LCDM-ext}
\begin{tabular}{lllc}
 \textbf{ Model } &  \textbf{ Parameter } &  \textbf{ prior range} &  \textbf{ posterior } \\
\hline
\hline
\textbf{$w$CDM}  & $w_0$ & -2.0 - 0.0 & $-0.873\pm0.077$\\
\hdashline
\textbf{$w_0 w_a$CDM} & $w_0$ &  -2.0 - 0.0 & $-0.943\pm0.168$\\
& $w_a$ &  -4.0 - 4.0 & $0.156\pm0.361$\\
\hdashline
\textbf{$o\Lambda$CDM} & $\Omega_k$ & -1.0 - 1.0 & $-0.0024\pm0.0032$\\ 
\hline
                    &$\beta_1$ & 0 - 2.0 & $1.23 \pm 0.29$ \\
                    &$\beta_2$ & 0 - 2.0 & $0.93 \pm 0.44$\\
\textbf{Scalar-tensor} & $\lambda_1^2 \times 10^{-6}
                    $    & 0 - $1$ & $0.49 \pm 0.29$\\
                    & $\lambda_2^2 \times 10^{-6}$    & 0 - $1$ & $0.41 \pm 0.28$\\
                    &  $s$ & 1.0 - 4.0 &   $2.80 \pm 0.84$\\
\hdashline
                    &$\beta_1$ & 1.0 - 2.0 & $ < 1.008$ \\
\textbf{Chameleon} & $B_0$    & 0 - 1.0 & $<1.0$\\
                    &  $s$ & 1.0 - 4.0 &  $2.27<s<4.0$ \\
\hdashline
                    &$\beta_1$ & 0 - 2.0 & $0.932\pm0.031$ \\
\textbf{eChameleon}        & $B_0$    & 0 - 1.0 & $<0.613$\\
                    &  $s$ & 1.0 - 4.0 &  $2.69<s<4.0$ \\
\hdashline                    
$\mathbf{f(R)}$  & $B_0$ & $10^{-10}$ - $10^{-4}$ & $<1.32 \times 10^{-5}$\\
\hdashline
\textbf{Growth index} & $\gamma$ & 0.2 - 0.8 & $0.611\pm0.072$\\
\hline
\end{tabular}
\end{center}
\end{table}

\begin{table*}[bp!]
\begin{center}
\caption{The list of standard $\Lambda$CDM parameters used in our analysis. For each parameter we provide its symbol, prior range, central value and $1\sigma$ error. We have used the same prior as Planck2013 on these parameters. We have also marginalized over all the nuisance parameters of Planck likelihood. We report the results for each of the model analyzed in this paper.}
\label{tbl:LCDM-par}
\begin{tabular}{lcccccc}
\textbf {Models} & $\mathbf{\Omega_b h^2}$ & $\mathbf{\Omega_c h^2}$ & $\mathbf{100\theta_{MC}}$ & $\mathbf{\tau}$ & $\mathbf{n_s}$ & $\mathbf{\ln(10^{10}A_s)}$ \\
\hline                                 
\textbf {prior range} &            0.005-0.10                 & 0.001-0.99                    & 0.50-10.0                     &  0.01-0.8                    & 0.9-1.1                       & 2.7-4.0        \\
%\textbf {$\Lambda$CDM} &   $0.0219 \pm 0.0003$  & $0.1199  \pm 0.0031$ & $1.0412  \pm 0.0007$ & $0.0649 \pm 0.0344$ & $0.9564 \pm 0.0096$ & $3.0402 \pm 0.0637$ \\
\textbf {$\Lambda$CDM} &   $0.0219 \pm 0.0002$  & $0.1208  \pm 0.0020$ & $1.0410  \pm 0.0006$ & $0.0442 \pm 0.0236$ & $0.953 \pm 0.0068$ & $3.0007 \pm 0.0450$ \\
\textbf {$w$CDM} &               $0.0221 \pm 0.0003$ & $0.1183  \pm 0.0028$ &  $1.0414  \pm 0.0006$ & $0.0911 \pm 0.0449$ & $0.9615 \pm 0.0097$ & $3.0884  \pm 0.0843$ \\
\textbf {$w_0 w_a$CDM} &   $0.0221 \pm 0.0003$ & $0.1181 \pm 0.0029$  & $1.0415 \pm 0.0007$   & $0.0906 \pm 0.0454$ & $0.9619 \pm 0.0099$ & $3.0871 \pm 0.0850$  \\
\textbf {$o\Lambda$CDM} &                    $0.0220 \pm 0.0003$  &$0.1191 \pm 0.0031$   & $1.0413 \pm 0.0007$  & $0.0518 \pm 0.0278$ & $0.9582 \pm 0.0094$  &  $3.0118 \pm 0.0514$   \\
\textbf {Scalar-tensor} &           $0.0221 \pm 0.0003$    & $0.1199\pm0.0020$   &  $1.0412 \pm 0.0006$    & $0.0333\pm0.0198$  & $0.9591 \pm 0.0071$ & $2.9769\pm0.0377$ \\
\textbf {Chameleon} &           $0.0219\pm0.0002$    & $0.1205\pm0.0020$   &  $1.0411\pm0.0006$    & $0.0390\pm  0.0222$  & $0.9539\pm 0.0067$ & $2.9894 \pm 0.0425$ \\
\textbf {eChameleon} &           $0.0218\pm0.0003$    & $0.1222\pm0.0023$   &  $1.0409\pm0.0006$    & $0.1313\pm0.0467$  & $0.9537\pm0.0079$ & $3.1780\pm0.0914$ \\
\textbf {$f(R)$} &                    $0.0221 \pm 0.0004$  & $0.1182 \pm 0.0033$ &  $1.0414 \pm 0.0008$  & $0.0733 \pm 0.0354$ & $0.9607 \pm 0.0101$ & $3.0526 \pm 0.0663$ \\
\textbf {growth index ($\gamma$) } & $0.0218\pm0.0003$   &  $0.1214\pm0.0023$   & $1.0409\pm0.0006$     & $0.0699\pm0.0400$  & $0.9525\pm0.0075$ & $3.0534\pm0.0788$ \\
\hline
\end{tabular}
\end{center}
\end{table*}

%\FloatBarrier

\begin{figure}
\includegraphics[width=0.5\textwidth]{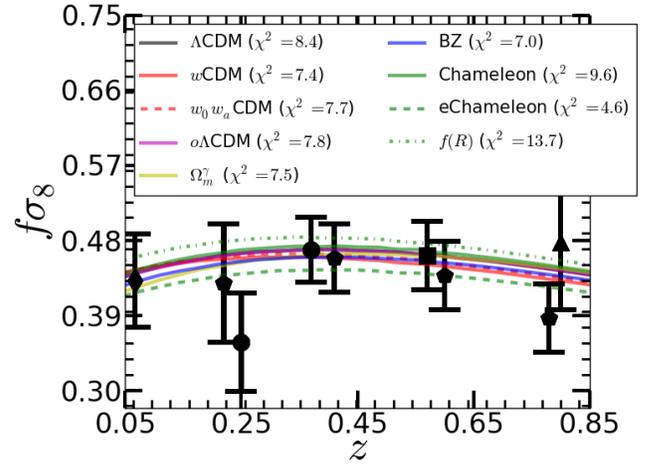}
\caption{The black points show the corrected $f\sigma_8$ used in our analysis, along with the errorbar. Lines of different colors show the best fit for the various models used in our analysis. The best fit and $\chi^2$ are for the case of Planck +$f\sigma_8$ + eCMASS fits. Notice that the eChameleon model predicts the smallest growth rate by preferring lower values of the coupling constant ($\beta_1$), even though the scalar amplitude of primordial power spectrum is high. }
\label{fig:fs8z_bestfit}
\end{figure}

%\begin{figure}
%\includegraphics[width=0.4\textwidth]{plots/Like-2D/All-sigma8-omegam_2D.png}
%\caption{The figure shows the $1\sigma$ and $2\sigma$ regions for each of the models considered in this paper in $\Omega_m$-$\sigma_8$ plane. It shows that the posterior likelihood is consistent for each of the model in this parameter space.}
%\label{fig:Ms8}
%\end{figure}

\begin{figure}
\includegraphics[width=0.4\textwidth]{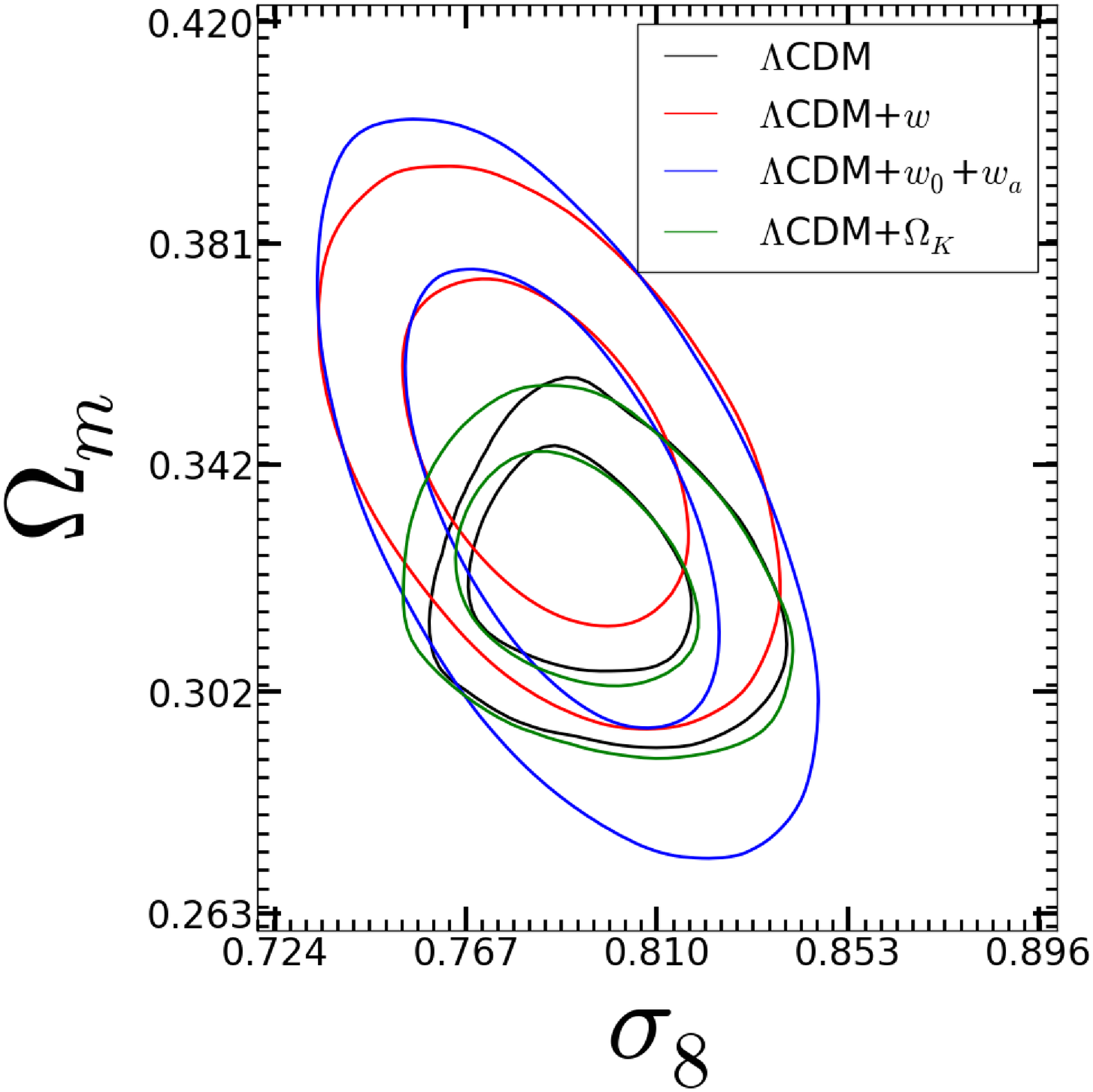} \\
\includegraphics[width=0.4\textwidth]{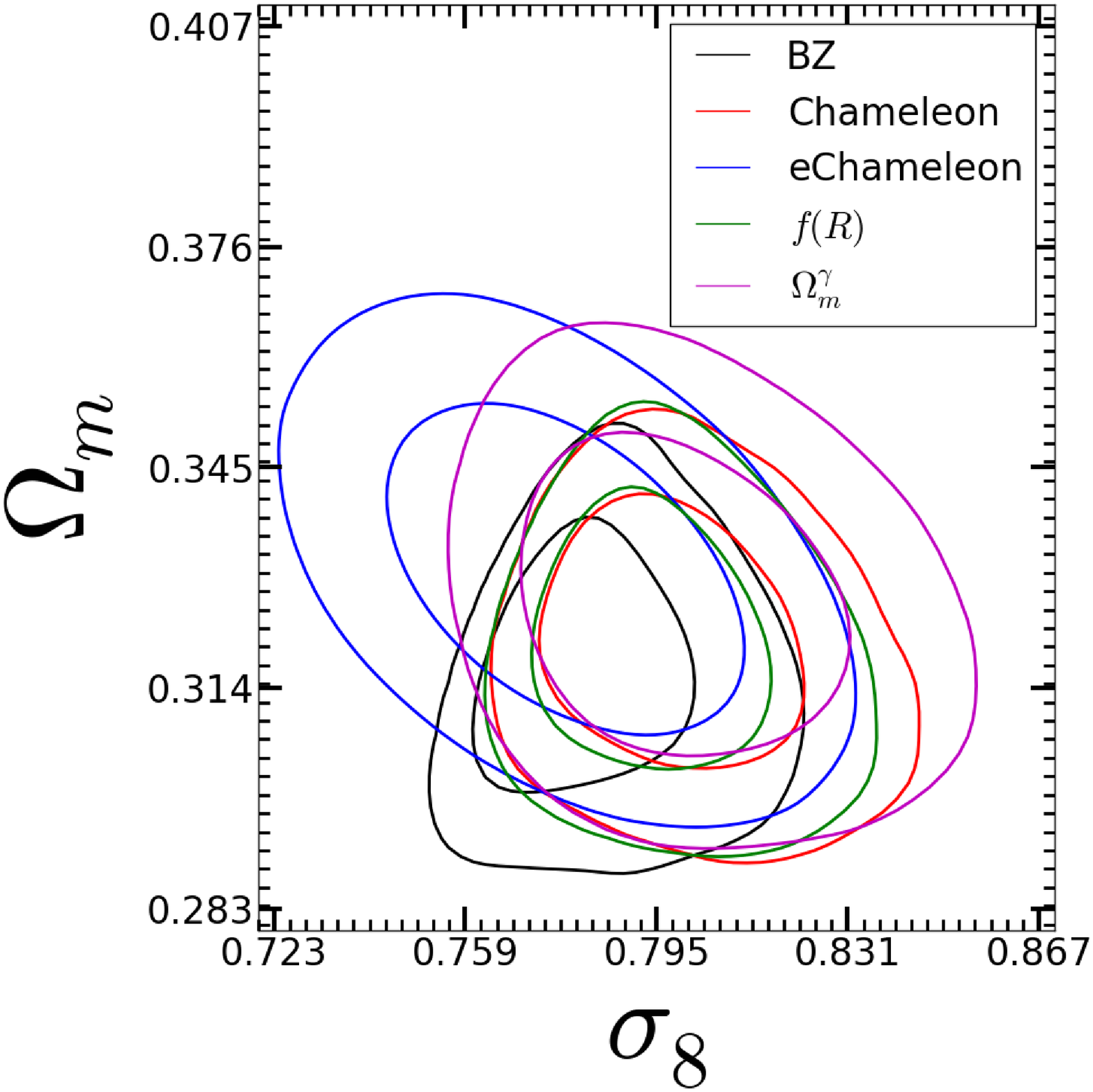}
\caption{The figure shows the $1\sigma$ and $2\sigma$ regions for each of the models considered in this paper in $\Omega_m$-$\sigma_8$ plane. It shows that the posterior likelihood is consistent for each of the model in this parameter space. The top plots shows the models which are extension to $\Lambda$CDM and the bottom plot shows the {\it modified gravity} models.}
\label{fig:Ms8}
\end{figure}

We have combined CMB data set and measurements of growth from various redshift surveys in order to constrain the parameters of standard cosmology ($\Lambda$CDM), extended cosmology models  and {\it modified gravity}. Our analysis gives consistent constraints for the standard $\Lambda$CDM parameters \{$\Omega_b h^2, \Omega_c h^2, 100\Theta_{MC} ,\tau, n_s, log(10^{10} A_s) $\} as shown in Table \ref{tbl:LCDM-par}. Figure \ref{fig:Ms8} shows the constraint on $\Omega_m$-$\sigma_8$ plane for $\Lambda$CDM, wCDM, $o\Lambda$CDM, Scalar-tensor model, Chameleon gravity, eChameleon, $f(R)$  and  growth index parametrization. Theses are our best constraints obtained using planck +eCMASS + $f\sigma_8(z)$. 

\subsection{$\Lambda$CDM}
\begin{figure} 
\includegraphics[width=0.15\textwidth]{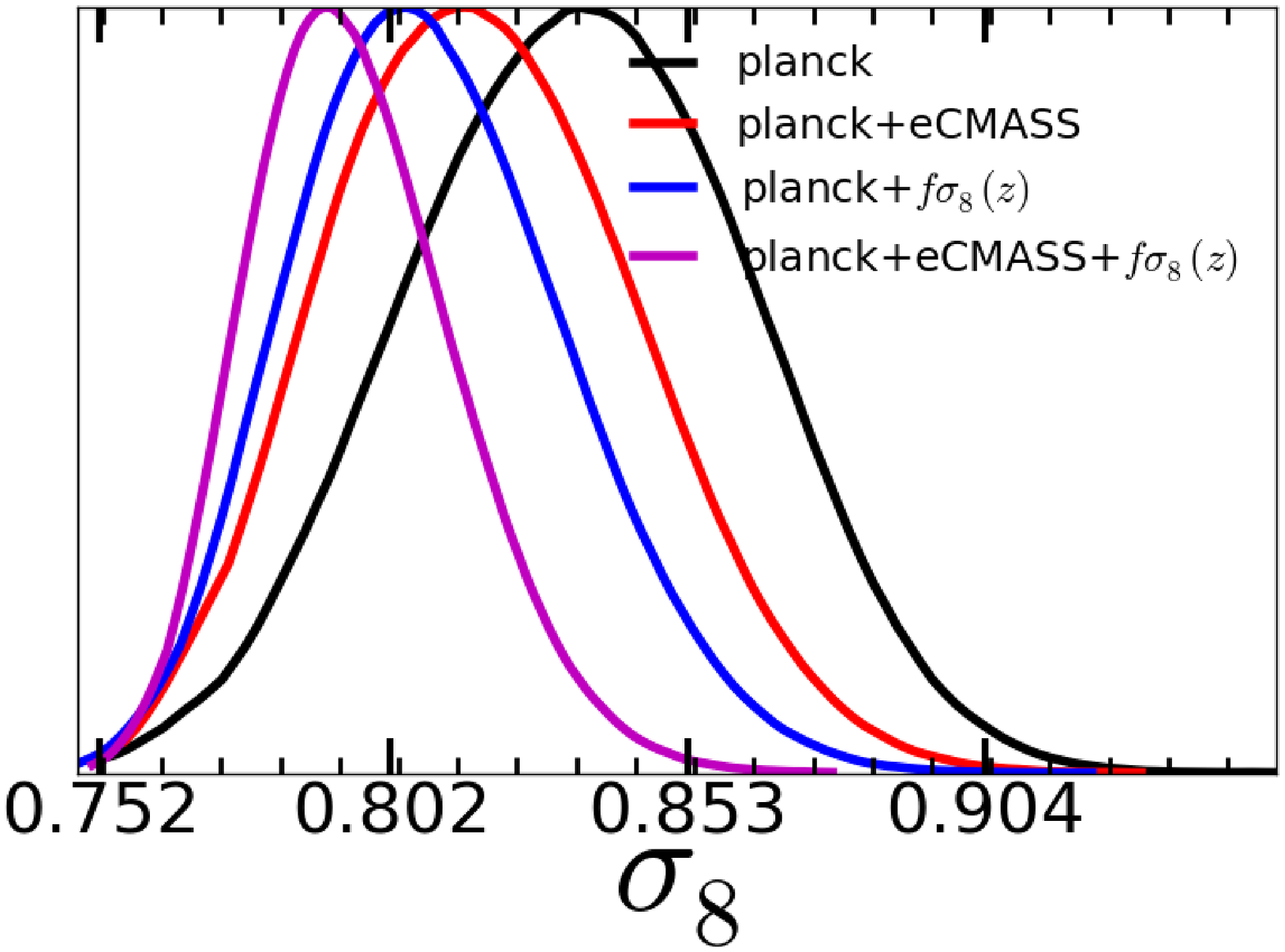}
\includegraphics[width=0.15\textwidth]{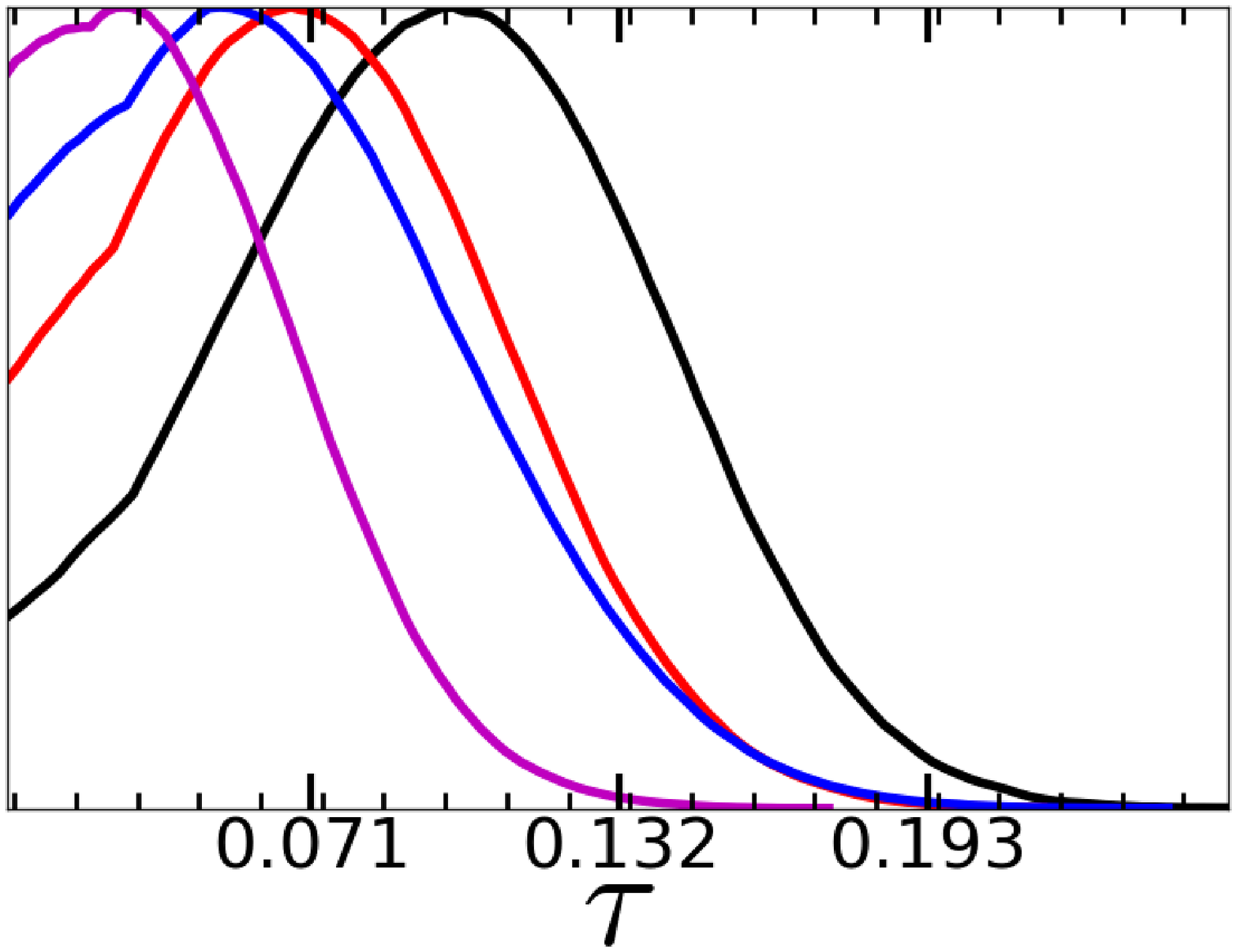}
\includegraphics[width=0.15\textwidth]{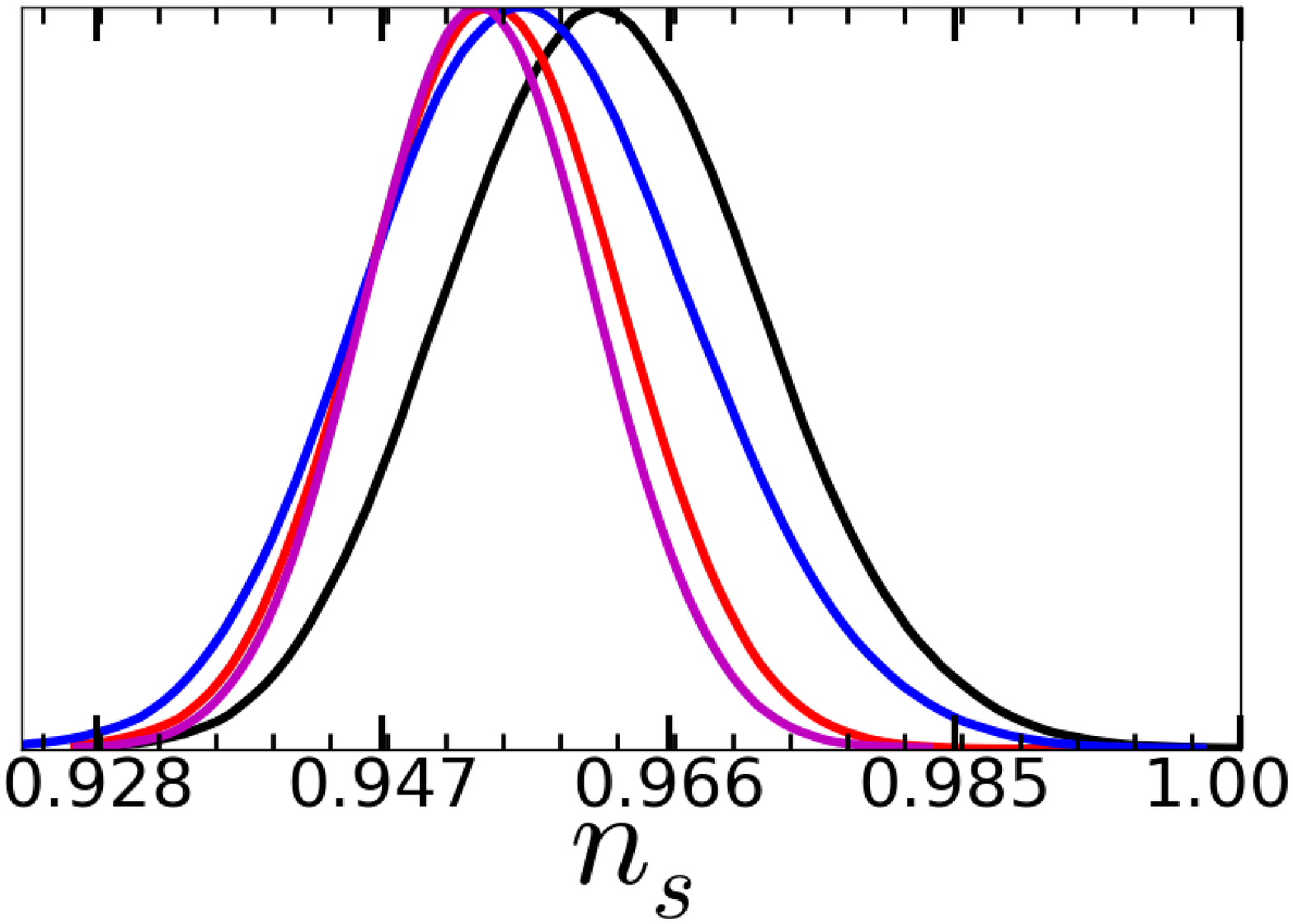}\\
\includegraphics[width=0.15\textwidth]{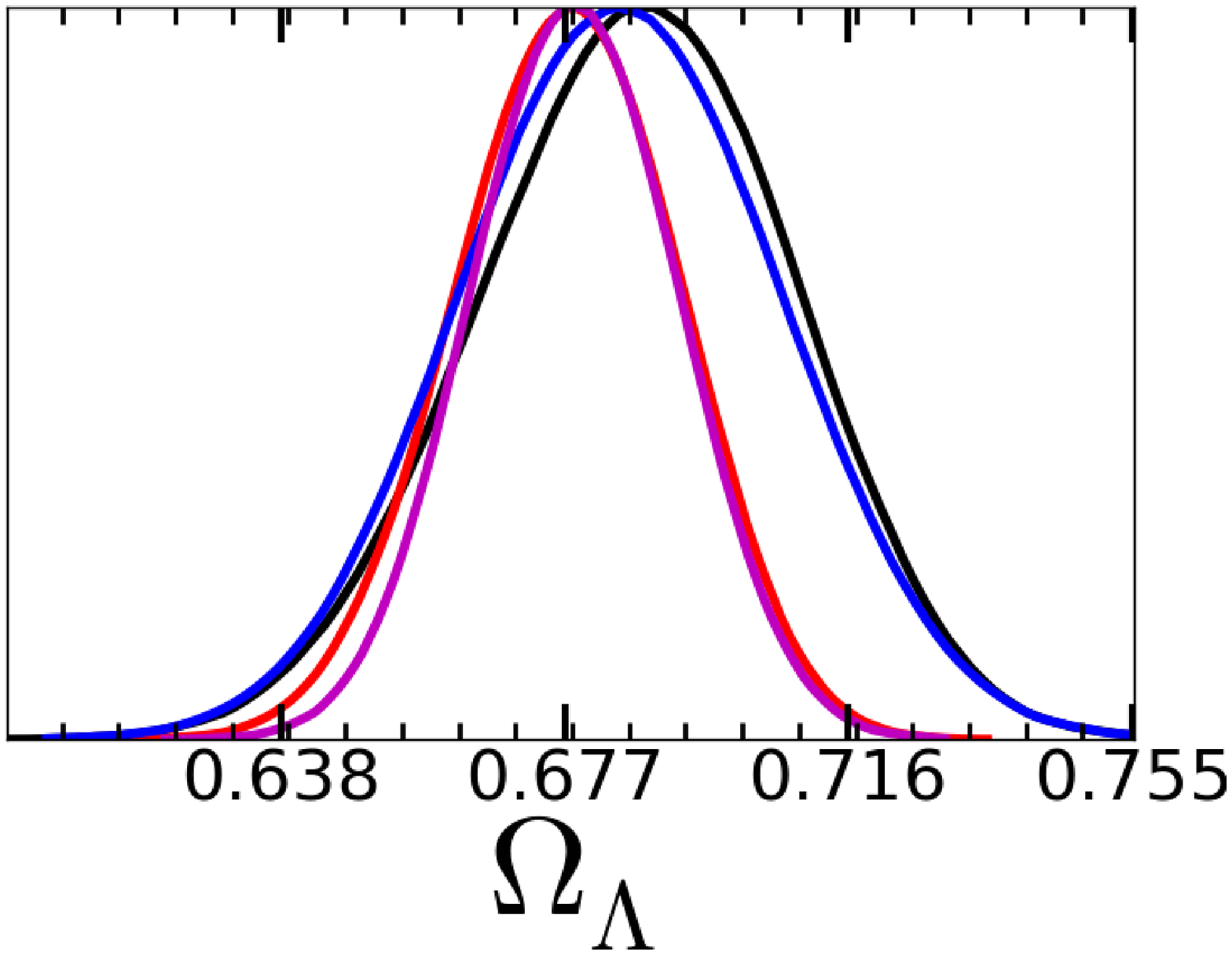}
\includegraphics[width=0.15\textwidth]{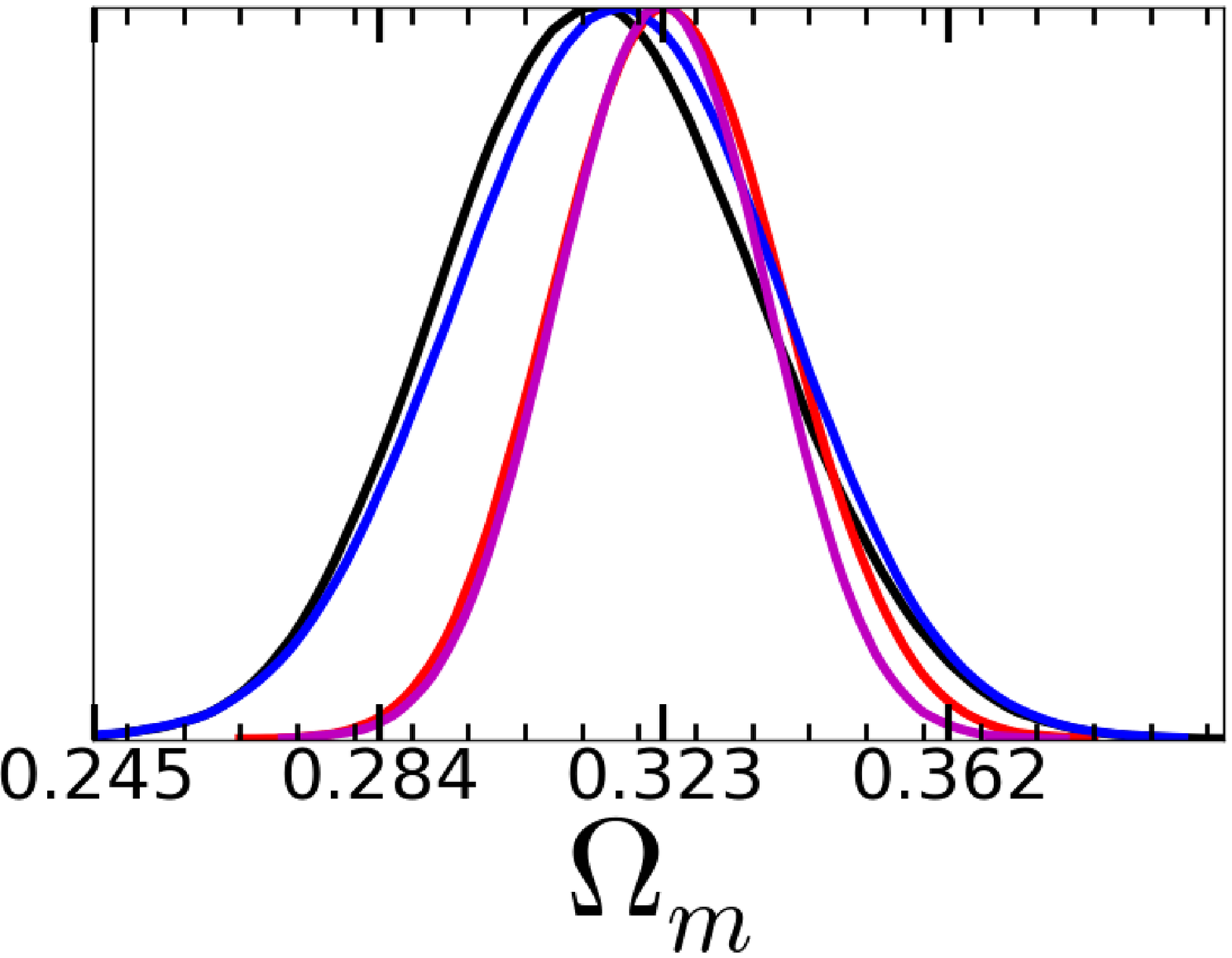}
\includegraphics[width=0.15\textwidth]{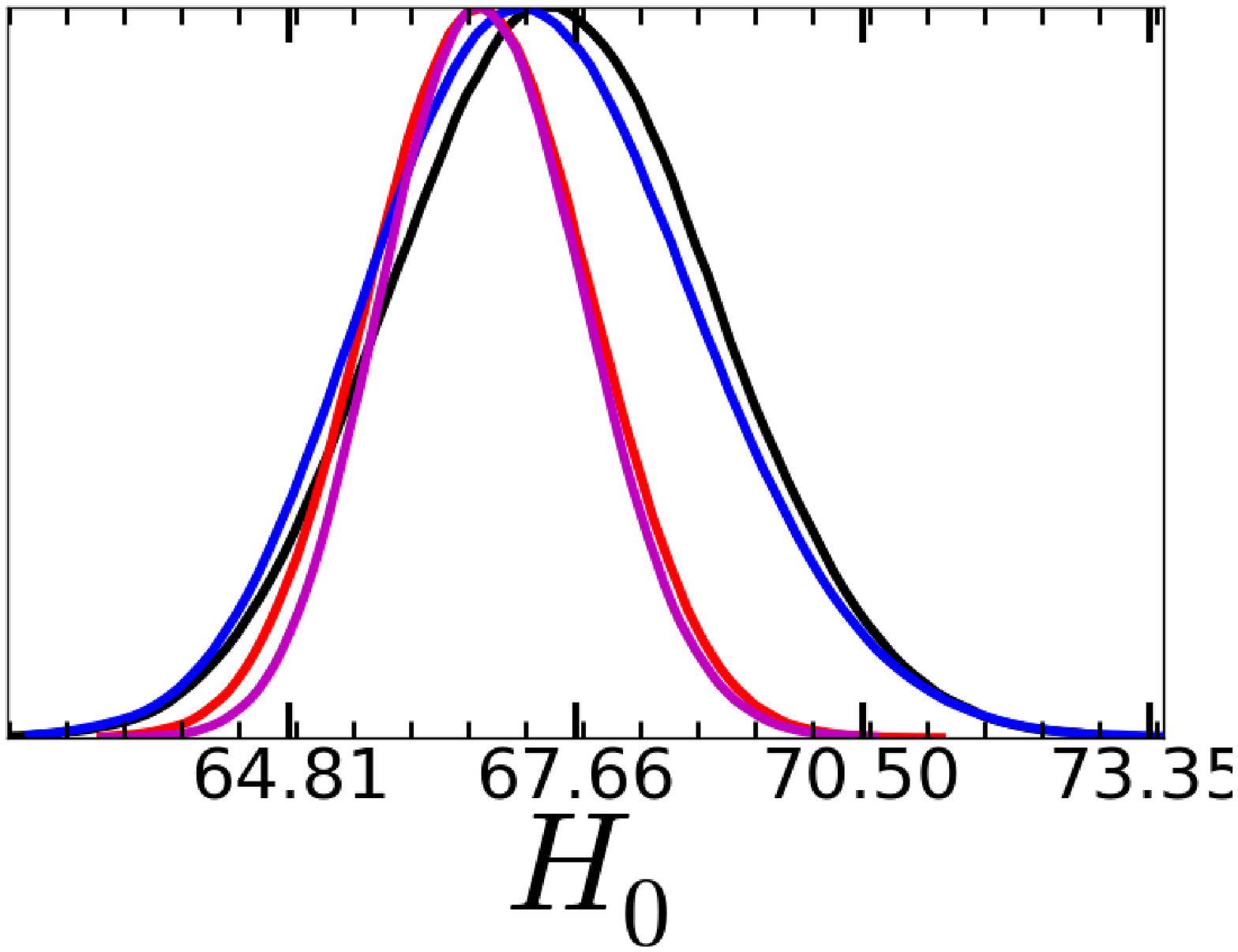}
\caption{ \textbf{$\Lambda$CDM:} We use GR as the model for gravity to determine the growth factor and fit for $f\sigma_8(z)$ and eCMASS measurement with Planck likelihood. The black line shows the contraints from Planck 2013 alone. The red ,blue and magenta lines are posterior obtained for the data set combinations planck+eCMASS, planck+$f\sigma_8(z)$ and planck+eCMASS+$f\sigma_8(z)$ respectively.  The two most prominent effect are in optical depth $\tau$ and scalar amplitude of primordial power spectrum $A_s$. Which is also reflected in the derived parameter $\sigma_8$ and mid redshift of re-ionization $z_{re}$.}
\label{fig:LCDM}
\end{figure}

Figure \ref{fig:LCDM} shows the one dimensional marginalized likelihood for standard $\Lambda$CDM cosmology. The black line shows the contraints from Planck 2013 alone. The red ,blue and magenta lines are posterior obtained for the data set combinations planck+eCMASS, planck+$f\sigma_8(z)$ and planck+eCMASS+$f\sigma_8(z)$ respectively. Our parameter constraints are completely consistent with the Planck 2013 results. Adding measurements of the growth rate to Planck data does not improve the results (see Figure \ref{fig:LCDM}) due to already tight constraints from Planck observations (see Figure \ref{fig:fs8z_data} ).

%We see \textcolor{red}{x} $\sigma$ deviation in $\sigma_8$ which points toward slight tension in the measurement of $\sigma_8$ between various measurement (add reference).

\subsection{Dark Energy Equation of state ($w$CDM)}
\begin{figure}
\includegraphics[width=0.4\textwidth]{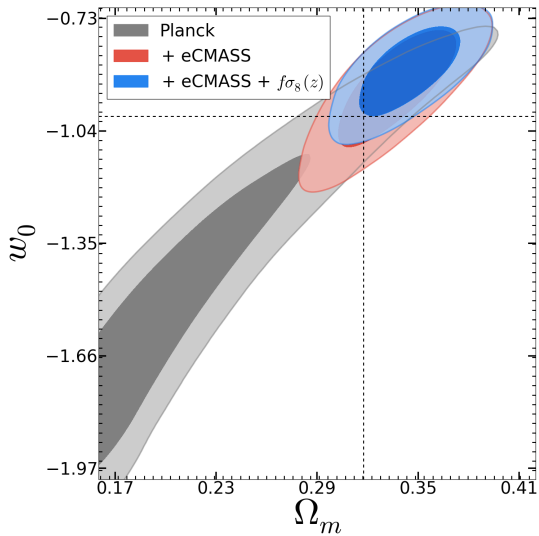}
\caption{\textbf{wCDM:} The two dimensional posterior likelihood $w$ and $\Omega_m$  for $w$CDM. The grey contour is for Planck ($w_0=-1.27\pm0.42$), red contour is combined constraint from Planck and eCMASS ($w_0=-0.92\pm0.10$). The blue contour represents constraint from combining Planck with eCMASS and $f\sigma_8(z)$ ($w_0=-0.87\pm0.077$).}
\label{fig:wCDM}
\end{figure}

We have looked at the $w$CDM, i.e the one parameter extension of $\Lambda$CDM where the dark energy equation of state is a constant, $w$.
Figure \ref{fig:wCDM} shows the two dimensional likelihood of $w_0$ and $\Omega_m$. The grey contours are Planck only constraint ($w_0=-1.27\pm0.42$), red contours are Planck and eCMASS ($w_0=-0.92\pm0.10$) and blue contours show Planck combined with eCMASS and growth factor measurements ($w_0=-0.87\pm0.077$). We obtain $w_0=-0.87\pm0.077$ (8.8\% measurement) which is consistent with the fiducial value of $w=-1$ for $\Lambda$CDM. The constraint we obtained is similar in precision as compared to BAO only, but has different degeneracy. Therefore combined measurement of growth rate and anisotropic BAO for all of these surveys will help us improve the precision of $w_0$. 

\subsection{Time-dependent Dark Energy ($w_0 w_a$CDM)}
\begin{figure}
\includegraphics[width=0.4\textwidth]{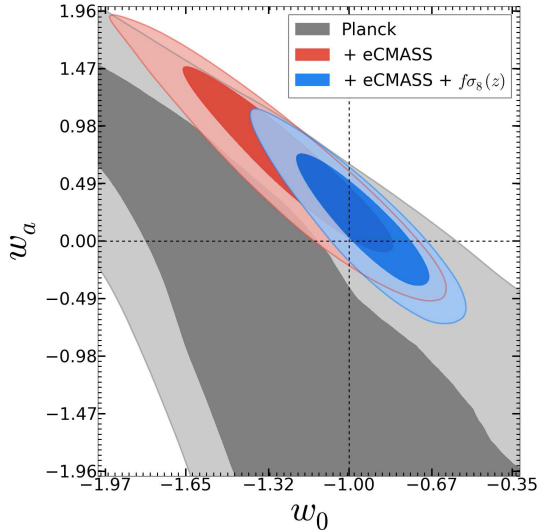}
\caption{\textbf{w0waCDM:} The two dimensional posterior likelihood of $w_0$ and $w_a$  for time-dependent dark energy model. The grey contour is for Planck ($w_0=-0.99\pm0.52, w_a=-1.50\pm1.46$), red contour is combined constraint from Planck and eCMASS ($w_0=-1.23\pm0.26, w_a=0.63\pm0.49$). The blue contour represents results from combining Planck with eCMASS and $f\sigma_8(z)$ ($w_0=-0.94\pm0.17,w_a=0.16\pm0.36$).}
\label{fig:w0waCDM}
\end{figure}

The $w$CDM model which proposes a constant dark energy is limited in its physical characteristics. Many models propose time-dependent dark energy which is popularly tested using linear relation $w(z)=w_0 +w_a \frac{z}{1+z}$ with $w_0$ and $w_a$ as free parameters. This model has been shown to match exact solutions of distance, Hubble, growth to the $10^{-3}$ level of accuracy \citep{Putter2008JCAP...10..042D} for a wide variety of scalar field (and {\it modified gravity}) models. The dynamical evolution of $w(z)$ can change the growth factor significantly and leave an imprint on the CMB. The combination of CMB and collection of growth factor at different redshifts is a unique way to test the time-dependent dark energy model.

Figure \ref{fig:w0waCDM} shows the $1\sigma$ and $2\sigma$ region for ($w_0,w_a$). The grey contour is from the Planck temperature power spectrum data alone ($w_0=-0.99\pm0.52, w_a=-1.50\pm1.46$). The red contours are from Planck and eCMASS ($w_0=-1.23\pm0.26, w_a=0.63\pm0.49$) and blue contour shows planck combined with eCMASS and growth factor measurement ($w_0=-0.94\pm0.17,w_a=0.16\pm0.36$).  The $\Lambda$CDM prediction of $(w_0,w_a)=(-1,0)$ is completely consistent with our posterior. We have obtained constraint on $w_0=-0.94\pm0.17$ (18\% measurement) and $1+w_a=1.16\pm0.36$ (31\% measurement) which is stronger constraint than the current best measurement of $w_a=-0.2\pm0.4$ from \citet{Aubourg2014}.

\subsection{Spatial Curvature ($o\Lambda$CDM)}
\begin{figure}
\includegraphics[width=0.4\textwidth]{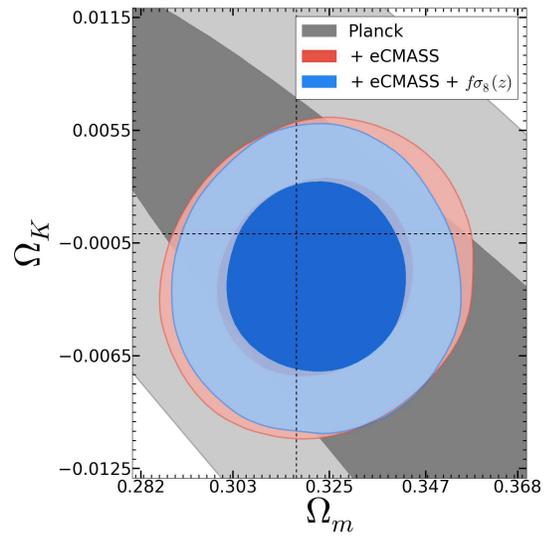}
\caption{\textbf{$o\Lambda$CDM:} The two dimensional posterior likelihood of $\Omega_k$ and $\Omega_m$  for $o\Lambda$CDM. The grey contour is for Planck ($\Omega_k=-0.060\pm0.047$), red contour is combined constraint from Planck and eCMASS ($\Omega_k=-0.0024\pm0.0034$). The blue contour represents results from combining Planck with eCMASS and $f\sigma_8(z)$ ($\Omega_k=-0.0024\pm0.0032$).}
\label{fig:kCDM}
\end{figure}

We consider a model with spatial curvature parametrized with $\Omega_K$ as free parameter called $o\Lambda$CDM along with $\Lambda$CDM parameters. Figure \ref{fig:kCDM} shows the posterior for the $\Omega_K$ and $\Omega_m$ plane. The grey contour is from the Planck temperature power spectrum data alone ($\Omega_k=-0.060\pm0.047$). The red contours are from Planck and eCMASS ($\Omega_k=-0.0024\pm0.0034$) and blue contour shows Planck combined with eCMASS and growth factor measurements ($\Omega_k=-0.0024\pm0.0032$).  The $\Lambda$CDM prediction of $\Omega_k=0$ is completely consistent with our posterior. We have obtained constraint on $1+\Omega_k=0.9976\pm0.0032$ (0.3\% measurement) which is competitive with the current best measurements \citep{Aubourg2014}. It will be interesting to see if combined RSD and BAO at all redshifts will give any improvement on the precision of curvature.

\subsection{Scalar-Tensor Gravity (BZ parametrization)}

%\begin{figure} 
%\includegraphics[width=0.23\textwidth]{plots/Like-1d/planck-RSD-BZ-B1.eps}
%\includegraphics[width=0.23\textwidth]{plots/Like-1d/planck-RSD-BZ-B2.eps}\\
%\includegraphics[width=0.15\textwidth]{plots/Like-1d/planck-RSD-BZ-lambda1_2.eps}
%\includegraphics[width=0.15\textwidth]{plots/Like-1d/planck-RSD-BZ-lambda2_2.eps}
%\includegraphics[width=0.15\textwidth]{plots/Like-1d/planck-RSD-BZ-ss.eps}\\
%\caption{ \textbf{BZ:} \textcolor{red}{ We use BZ parametrization of Scalar-tensor theories as the model for gravity to determine the growth factor and fit for $f\sigma_8(z)$ and eCMASS measurement with planck likelihood. The black, red ,blue and magenta lines are posterior obtained for the data set combinations planck+eCMASS, planck+$f\sigma_8(z)$ with fixed $k$, planck+$f\sigma_8(z)$ with averaged over $k$ and planck+eCMASS+$f\sigma_8(z)$ respectively.  Table \ref{tbl:LCDM-par} provides the constraints on basic $\Lambda$CDM parameters and Table \ref{tbl:LCDM-ext} provides the constraint on five extension parameters $(\beta_1,\beta_2,\lambda_1^2,\lambda_2^2,s)$. }}
%\label{fig:BZ-1D}
%\end{figure}

\begin{figure}
\includegraphics[width=0.4\textwidth]{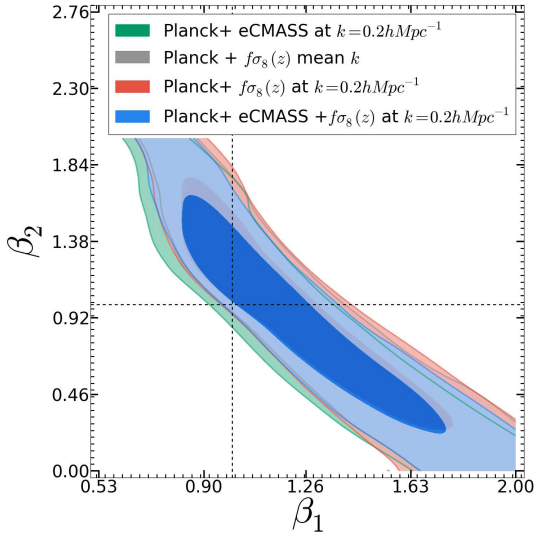}
\caption{\textbf{BZ:}  The two dimensional posterior likelihood of $\beta_1$-$\beta_2$  for five parameter Scalar-tensor theory parametrized through the BZ form of~(\ref{mu_gamma_BZ}). The green contour is the combined constraint from Planck and eCMASS ($\beta_1=1.18\pm0.29, \beta_2=0.95\pm0.43$).The grey contour is the combined constraint from Planck and $f\sigma_8(z)$ with averaged over k ($\beta_1=1.24\pm0.3,\beta_2=0.96\pm0.45$), red contour is the combined constraint from Planck and $f\sigma_8(z)$ at effective $k=0.2h$ Mpc$^{-1}$ ($\beta_1=1.24\pm0.3,\beta_2=0.95\pm0.45$). The blue contour represents results from the combination of Planck, $f\sigma_8(z)$ and eCMASS ($\beta_1=1.23\pm0.29,\beta_2=0.93\pm0.44$). }
\label{fig:BZ-2D}
\end{figure}

The general Scalar-tensor theories of gravity is analyzed using five parameter model called BZ parametrization  . The five parameters of Scalar-tensor gravity ($\beta_1, \beta_2,\lambda_1,\lambda_2, s$) are constrained along with the standard $\Lambda$CDM parameters using Planck, $f\sigma_8(z)$ and eCMASS measurements. BZ model predicts a scale dependent growth rate ($f\sigma_8(k,z)$), whereas the measurements are at some effective $k$. In order to incorporate the k-dependence in our analysis, we use the two different approaches described in Section \ref{scale_dep}. Figure \ref{fig:BZ-2D} shows two dimensional posterior in the plane ($\beta_1,\beta_2$) . The green contour is combined constraint from Planck and eCMASS ($\beta_1=1.18\pm0.29, \beta_2=0.95\pm0.43$). The grey contour is the combined constraint from Planck and $f\sigma_8(z)$ with averaged over k ($\beta_1=1.24\pm0.3,\beta_2=0.96\pm0.45$); the red contour is the combined constraint from Planck and $f\sigma_8(z)$ at effective $k=0.2h$ Mpc$^{-1}$ ($\beta_1=1.24\pm0.3,\beta_2=0.95\pm0.45$). The blue contour represents results from the combination of Planck, $f\sigma_8(z)$ and eCMASS ($\beta_1=1.23\pm0.29,\beta_2=0.93\pm0.44$). We obtain the following joint constraint on the five BZ  parameters: $\beta_1=1.23\pm0.29$, $\beta_2=0.93\pm0.44$, $\lambda_1^2(\times 10^{-6})=0.49\pm0.29$ ,$\lambda_2^2(\times 10^{-6})=0.41\pm0.28$ and $s=2.80\pm0.84$. By looking at the joint 2D likelihood for $(\beta_1,\beta_2)$ in Fig.~\ref{fig:BZ-2D}, we notice that there is a strong degeneracy between the two parameters which reflects the degeneracy between $\mu$ and $\gamma$ for the observables that we are using. Similar results have been found in~\citep{Hojjati2012,Planck2015DEMG}. For the next models that we will discuss, $\beta_1$ and $\beta_2$ are not independent and this will allow data to place more stringent constraints. 

While the constraints on the length scale of the scalar field ($\lambda_1,\lambda_2$) and ($s$) are very broad, the one on the coupling, $\beta_1$ and $\beta_2$,  is the first ever constraint obtained on these parameters for general Scalar-tensor gravity. The discrepancy in the strength of the constraints on the coupling and on the length scale, can be linked to the fact that data strongly prefer values of the coupling constants close to 1. For such values,  the scale and time dependences in $(\mu,\gamma)$ becomes less important and therefore are loosely constrained. We will encounter this again in the Chameleon and $f(R)$ gravity cases.

\subsection{Chameleon Gravity}
\begin{figure}
\includegraphics[width=0.15\textwidth]{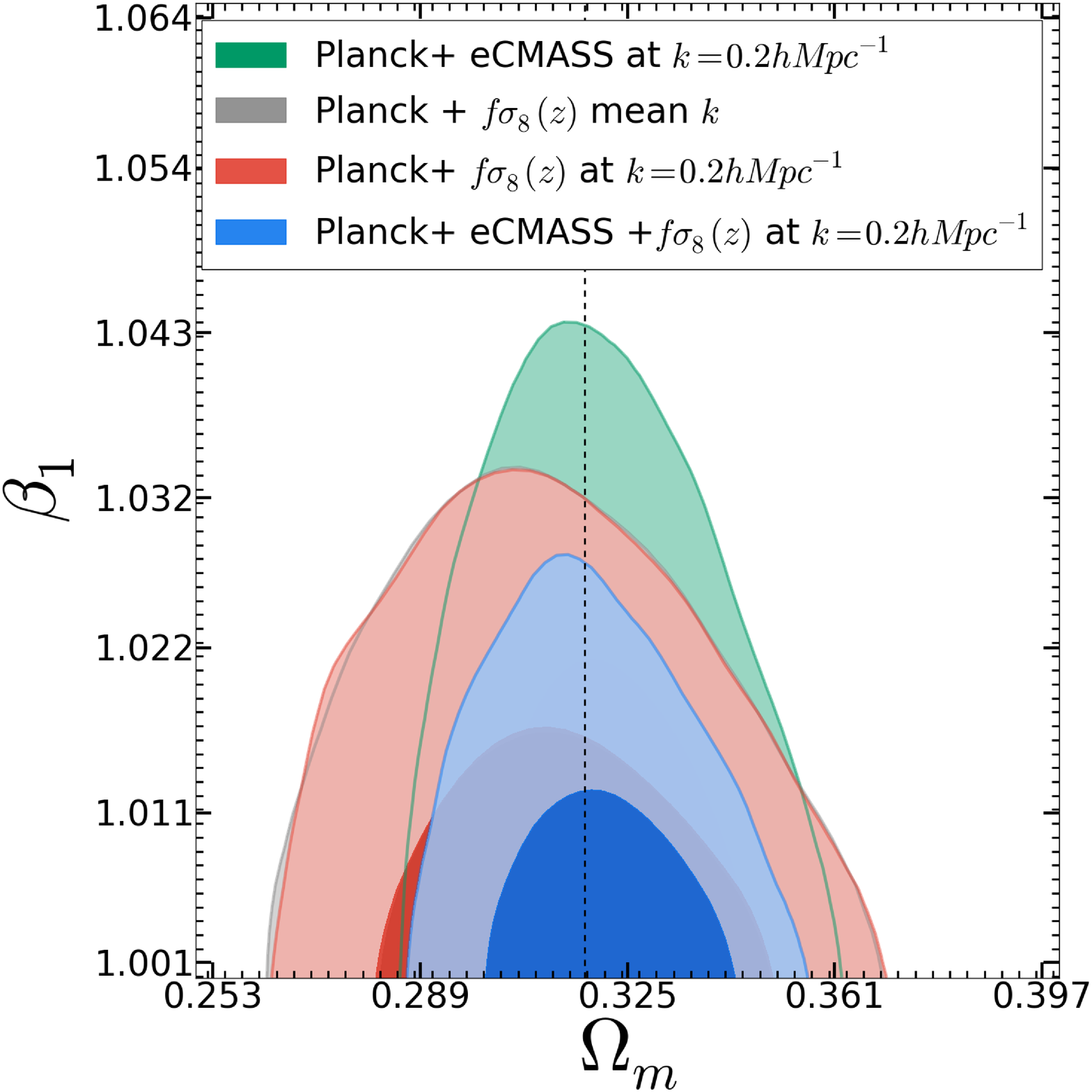}
\includegraphics[width=0.15\textwidth]{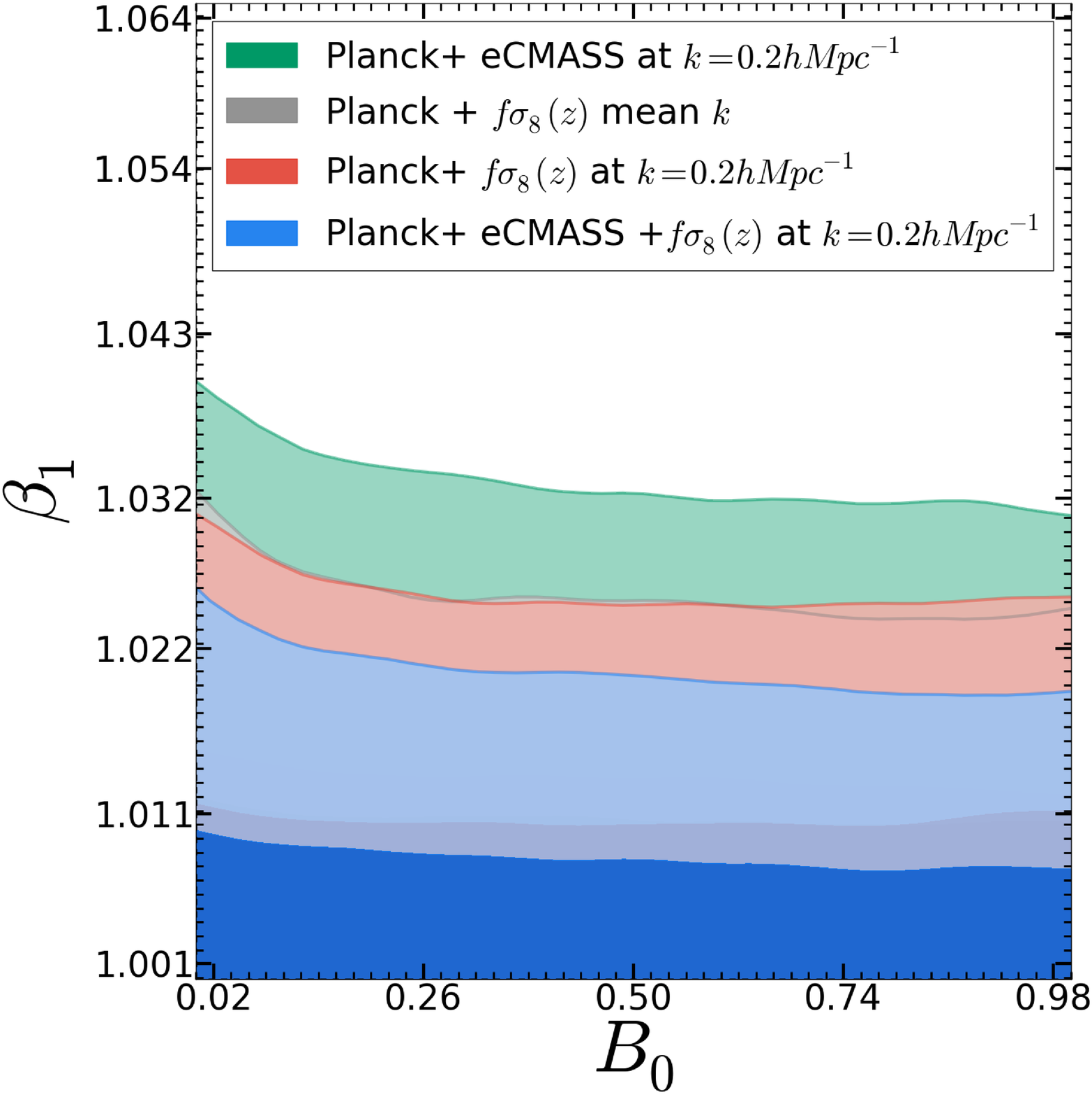}
\includegraphics[width=0.15\textwidth]{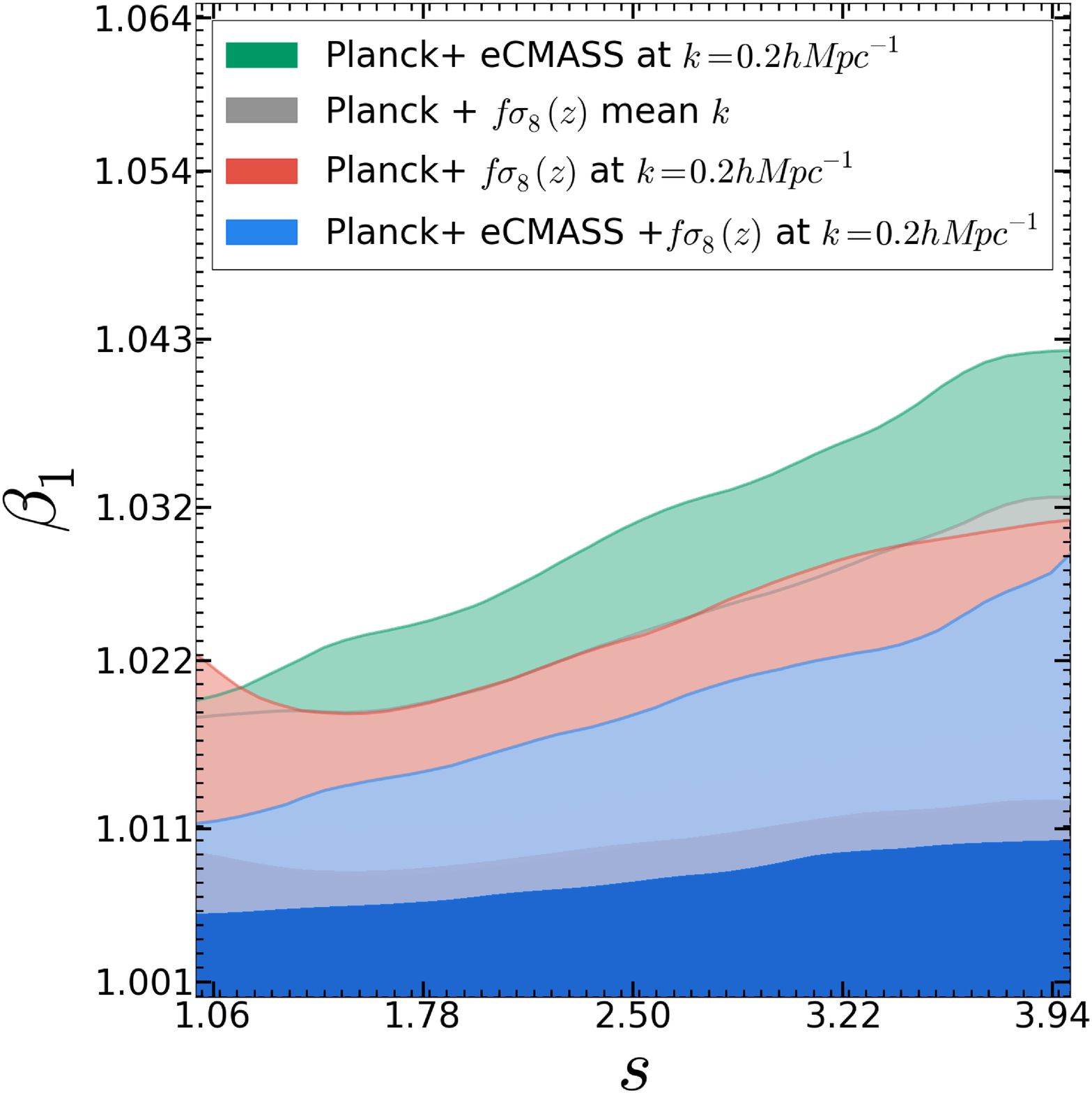}
\caption{\textbf{Chameleon Theory:} The two dimensional posterior likelihood for Chameleon gravity. The green contour is the combined constraint from Planck and eCMASS ($\beta_1<1.013$).The grey contour is the combined constraint from Planck and $f\sigma_8(z)$ with averaged over k ($\beta_1<1.010$), the red contour is the combined constraint from Planck and $f\sigma_8(z)$ at effective $k=0.2h$ Mpc$^{-1}$ ($\beta_1<1.010$). The blue contour represents results from the combination of Planck, eCMASS  and $f\sigma_8(z)$  ($\beta_1<1.008$).}
\label{fig:vCham-2D}
\end{figure}

\begin{figure}        
\includegraphics[width=0.4\textwidth]{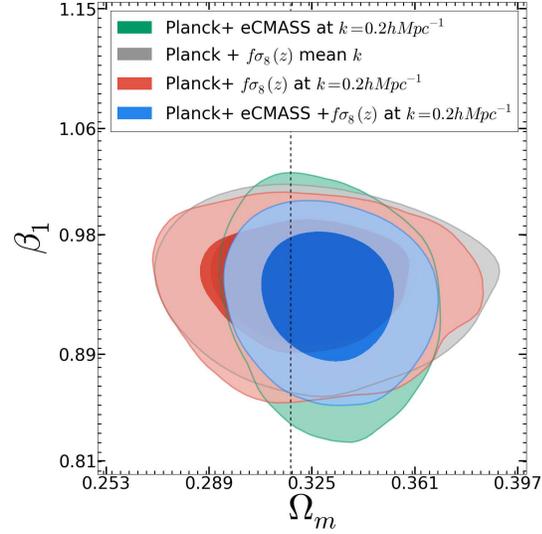}
\caption{\textbf{eChameleon Theory:}  The two dimensional posterior likelihood of $\beta_1$ and $\Omega_m$  for extended Chameleon gravity. The green contour is the combined constraint from Planck and eCMASS ($\beta_1=0.932\pm0.04$). The grey contour is the combined constraint from Planck and $f\sigma_8(z)$ with averaged over k ($\beta_1=0.940\pm0.032$), red contour is combined constraint from Planck and $f\sigma_8(z)$ at effective $k=0.2h$ Mpc$^{-1}$ ($\beta_1=0.936\pm0.032$). The blue contour represents results from the combination of Planck, $f\sigma_8(z)$ and eCMASS ($\beta_1=0.932\pm0.031$).}
\label{fig:Chambeta}
\end{figure}

The three parameters of Chameleon gravity ($\beta_1, B_0, s$) are constrained along with the standard $\Lambda$CDM parameters using Planck, $f\sigma_8(z)$ and eCMASS measurements. Chameleon models predict a scale dependent growth rate ($f\sigma_8(k,z)$), whereas the measurements are at some effective $k$. In order to incorporate the k-dependence in our analysis, we use the two different approaches described in Section \ref{scale_dep}.   Figure \ref{fig:vCham-2D} shows the two dimensional posterior in the plane ($\Omega_m,\beta_1$), ($B_0,\beta_1$) and ($s,\beta_1$). The grey and red contours show the posteriors from combined data set of Planck and growth rate measurements. The red contours are likelihood while evaluating the growth rate at an effective $k$ ($\beta_1<1.010$), whereas grey contours are for the case when we use an effective growth rate, averaged over the scales used in the actual $f\sigma_8$ measurement ($\beta_1<1.010$). The green contour is combined constraint from Planck and eCMASS ($\beta_1<1.013$). Finally, the blue contours show the posterior from combined data of Planck, eCMASS and growth rate ($\beta_1<1.008$). We obtain the following joint constraint on the three Chameleon parameters: $\beta_1<1.008$, $B_0<1.0$  and $2.27<s<4$. While the constraints on the length scale of the scalar field, $B_0$ and $s$ is very broad, the one on the coupling, $\beta_1$,  is very strong and predicts $\beta_1=1$ to 0.8\%, bringing $\mu$ to its GR value.
As we already discussed for the Scalar-tensor case, the discrepancy in the strenght of these constraints is due to the fact that data prefer values of the coupling constant close to 1, for which the time and scale dependence of $(\mu,\gamma_{slip})$ become negligible. This is even more the case for Chameleon models, where the theoretical prior forces $\beta_1 >1$, which corresponds to enhanced growth, and data consequently require very small values for this coupling, pushing $\mu$ very close to its GR value.

We have also looked at the extended Chameleon model where we allow $\beta_1$ to be less than 1 following previous analysis of this model. Figure \ref{fig:Chambeta} shows the two dimensional posterior in the plane ($\Omega_m,\beta_1$). The red contours are likelihood while evaluating the growth rate at an effective $k$ ($\beta_1=0.940\pm0.032$), whereas grey contours are for the case when we use an effective growth rate, averaged over the scales used in the actual $f\sigma_8$ measurement ($\beta_1=0.936\pm0.032$). The green contour is combined constraint from Planck and eCMASS ($\beta_1=0.932\pm0.04$). Finally, the blue contours show the posterior from combined data of Planck, eCMASS and growth rate ($\beta_1=0.932\pm0.031$). We obtain the following joint constraint on the three eChameleon parameters: $\beta_1=0.932\pm0.031$, $B_0<0.613$  and $2.69<s<4$. Like in the more general Scalar-tensor case, while the constraints on the length scale of the scalar field, $B_0$ and $s$ are very broad, the one on the coupling, $\beta_1$,  is an huge improvement on the previous constraint of $\beta_1=1.3\pm0.25$ (19.2 \% measurement) using WMAP CMB, SNe and ISW dataset \citep{Hojjati2011}. Let us notice that when we constrain jointly the three eChameleon parameters, data select a region in the parameter space which corresponds to $\beta_1<1$, i.e. to suppressed growth. This region excludes standard Chameleon models, including $f(R)$ theories, for which $\beta_1>1$ and the growth is enhanced. After all, as we have seen above and will see in the next Section, the same data place very stringent constraints on Chameleon and $f(R)$ models, forcing them to be very close to $\Lambda$CDM (see Figure \ref{fig:Chambeta} and \ref{fig:fR}). Hence the combination of data sets that we employ favor models with a suppressed growth rate, which adopting the BZ parametrization can be obtained with $\beta_1<1$; a suppressed growth was favored also by the data set used in~\citep{Planck2015DEMG}, although in that case the authors employed a time-dependent parametrization. Theoretically viable scalar-tensor models with a suppressed growth are discussed in~\citep{Perenon2015}, where they are analyzed via a scale-independent parametrization in the effective field theory language.
 
\subsection{$f(R)$ theory}
\begin{figure}
    \begin{center}        
        \includegraphics[width=0.4\textwidth]{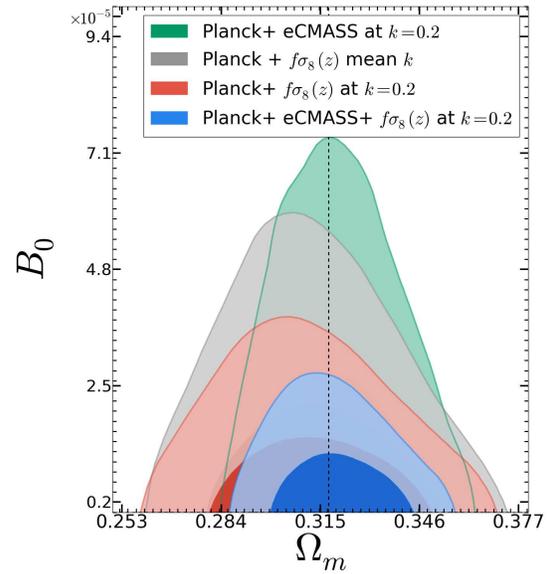}
     \end{center}
     \caption{\textbf{$f(R)$ Gravity:} The two dimensional posterior likelihood of $B_0$ and $\Omega_m$  for $f(R)$ gravity. The green contour is combined constraint from Planck and eCMASS ($B_0<3.43 \times 10^{-5}$). The grey contour is the combined constraint from Planck and $f\sigma_8(z)$ with averaged over k ($B_0<2.77 \times 10^{-5}$), the red contour is the combined constraint from Planck and $f\sigma_8(z)$ at effective $k=0.2h$ Mpc$^{-1}$ ($B_0<1.89 \times 10^{-5}$). The blue contour represents results from the combination of Planck, $f\sigma_8(z)$ and eCMASS ($B_0<1.36 \times 10^{-5}$)}
     \label{fig:fR}
\end{figure}

We consider one parameter ($B_0$) model of $f(R)$ gravity. The parameter $B_0$ parameterizes the deviation from $\Lambda$CDM. The model approaches GR when $B_0$ is zero.  Similar to Chameleon theory, $f(R)$ gravity predicts a scale dependent growth rate ($f\sigma_8(k,z)$). Figure \ref{fig:fR} shows the two dimensional posterior in $B_0$ and $\Omega_m$ plane. The green contour is combined constraint from Planck and eCMASS ($B_0<3.43 \times 10^{-5}$). The grey and red contours show posterior from combined data set of Planck and growth rate measurements. The red contours are likelihood while evaluating the growth rate at an effective $k$ ($B_0<1.89\times10^{-5}$) where as grey contours are for the case when we use effective growth rate, which is averaged over scales used in the actual $f\sigma_8$ measurements ($B_0<2.77\times10^{-5}$). The blue contours show the posterior from combined data of Planck, eCMASS and growth rate ($B_0<1.36\times10^{-5}$). We obtained $B_0<1.36\times10^{-5}$ ($1\sigma$ C.L.), which is an improvement by a factor of 4 on the most recent constraint from large scale structure of $B_0=5.7\times10^{-5}$ ($1\sigma$ C.L.) \citep{XU2015}. Our constraint is competitive with the constraint from solar system tests and clusters \citep{Hu2007,Schmidt2009}.

\subsection{Growth index ($\gamma$) parametrization}
\begin{figure}
    \begin{center}        
        \includegraphics[width=0.4\textwidth]{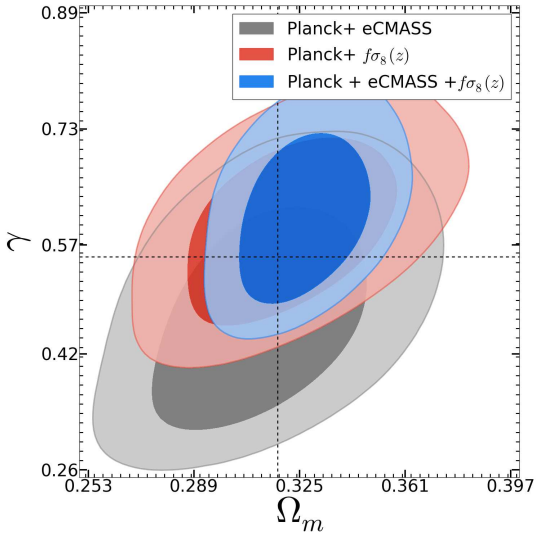}
     \end{center}
     \caption{\textbf{Growth index ($\gamma$):} The two dimensional posterior likelihood of $\gamma$ and $\Omega_m$  for growth index parametrization. The grey contour is the combined constraint from Planck and eCMASS ($\gamma = 0.477 \pm 0.096 $). The red contour is the combined constraint from Planck and $f\sigma_8(z)$ ($\gamma=0.595\pm0.079$). The blue contour represents results from the combination of Planck, $f\sigma_8(z)$ and eCMASS ($\gamma=0.612\pm0.072$)}
     \label{fig:Linder}
\end{figure}

The standard cosmological model, based on GR,  predicts a precise value for the growth factor in the linear regime, i.e.  $f=\Omega_m^{0.55}$.  In order to test  deviations from GR, we have parameterized the growth factor using growth index $\gamma$ ( cite Linder \& Cahn 2008) as $f=\Omega_m^{\gamma}$. The marginalized two dimensional likelihood for $\Omega_m$ and $\gamma$ is shown in Figure \ref{fig:Linder}. The grey contour is combined constraint from Planck and eCMASS ($\gamma=0.477\pm0.096$). The red contours show the constraint obtained using Planck and $f\sigma_8(z)$ measurement($\gamma=0.595\pm0.079$) and the blue contours are for combined data set of Planck with $f\sigma_8(z)$ and eCMASS ($\gamma=0.612\pm0.072$). We have obtained $\gamma = 0.612\pm0.072$ (11.7\% measurement) completely consistent with the general relativity prediction.

%% file: tex/discussion.tex
\section{Discussion}
\label{sec:discussion}

We have constrained the parameters of the standard cosmological model, $\Lambda$CDM, as well as those of  various extensions using the current measurements of growth rate between redshift 0.06 and 0.83 (Figure \ref{fig:Ms8}), eCMASS and Planck 2013. We have been careful with several important details while combining results from various surveys and different cosmologies of measurements. We have first showed that the standard $\Lambda$CDM parameter space has a consistent posterior, independent of the model considered except for Chameleon gravity. Next, we focused on each model and analyzed the constraint on its extension parameters. 
As for the standard model, $\Lambda$CDM, using the growth factor we do not improve constraints on any of its parameters because the growth rate is already highly constrained with Planck measurement for the standard model of cosmology. It is impressive to notice that $\Lambda$CDM, without any extra parameter, is completely consistent with the measurements of $f\sigma_8$ from very different galaxy types and redshifts. In the case of the extension where the dark energy equation of state is constant but free to vary, $w$CDM, we obtain $w_0=-0.87\pm0.077$ (8.8\% measurement). This is a 3.7 times improvement on the precision compared to Planck only measurement $w=-1.27\pm0.42$ (33\% measurement) and comparable to the 8\% measurement of \citet{Samushia2012}. Our measurement prefers $w<-1$ at $1\sigma$ level. We have also noticed that the growth rate and BAO have slightly different degeneracy for $w$CDM. This shows the potential to improve the constraint on $w$ by combining the growth rate and BAO measurements from a range of galaxy redshift surveys. However, one difficulty in doing so, is to model the correlation between the measurement of growth rate and BAO.

We also report one of the best measurements on the parameters of the model with a time-dependent equation of state, $w_0 w_a$CDM. We have measured 
$w_0=-0.94\pm0.17$ (18\% measurement) and $1+w_a=1.16\pm0.36$ (31\% measurement). This represents a significant improvement on $w_a$ compared to all other measurements \citep{Planck2013, Aubourg2014}. The measurements of $f\sigma_8$, $H$ and $D_A$ in eCMASS help to constrain $\Lambda$CDM parameters, while the evolution of the growth rate over a large redshift range, obtained through measurements of $f\sigma_8(z)$ at multiple redshifts,  improves the constrain on evolving dark energy. This hints at the potential of using combined growth rate and anisotropic BAO as function of redshifts, when future surveys like eBOSS and Euclid \citep{EuclidDef2011} will provide much stronger growth rate and BAO constraints at much higher redshifts. We have also looked at the possibility of a non-zero curvature for the universe,  $o\Lambda$CDM, finding  $1+\Omega_k=0.9976\pm0.0032$ (0.3\% measurement), which is same as the best constraint reported in \citep{Samushia2012}. We notice that the optical depth ($\tau$) and amplitude of scalar power spectrum ($A_s$) is relatively low for $o\Lambda$CDM, which predicts smaller redshift of reionization ($z_{re}=7.20 \pm 2.81$) but it is above the lower limit observed through Lyman$\alpha$ Forest observations (\cite{Becker2001GPlimit}).
 
We have also looked at some of the popular modifications of gravity and found  no significant deviations from GR using growth rate and Planck 2013 measurement. We have investigated general scalar-tensor theories under the  parametrization introduced in~\citep{Bertschinger2008}, constraining the corresponding five parameters. We have found constraints on the two coupling parameters ($\beta_1=1.23\pm0.29$, $\beta_2=0.93\pm0.44$) while the posterior of other three parameters were largely non-constraining. We then restricted to the subset of Chameleon theories, for which only three parameters are needed. While imposing a theoretical bound of $\beta_1>1$, we constrained the coupling of Chameleon theories to $\beta_1<1.008$ (1 $\sigma$ C.L.), while jointly varying the remaining two free parameters that describe the lengthscale of the scalar degree of freedom, $\{B_0,s\}$. While the latter are loosely constrained by data,  the constraint on the coupling is quite stringent. We explored also an extension of Chameleon models, that we dubbed eChameleon, where we let the coupling $\beta_1$ vary within the range $[0,2]$. Also in this case, data place a stringent bound on the coupling, while loosely constraining $\{B_0,s\}$. Interestingly, for this case data select a region where $\beta_1<1$, with the bound $\beta_1=0.932\pm0.031$; the latter corresponds to a region of the parameter space for which growth is suppressed. This improves significantly over previous analysis, e.g. the bound $\beta_1=1.3\pm0.25$ (19.2 \% measurement) obtained in  \citep{Hojjati2011} using WMAP CMB, SNe and ISW dataset. This excludes standard Chameleon models, including $f(R)$ theories, for which $\beta_1>1$ and the growth is enhanced. After all,  the same data place very stringent constraints on the latter models, forcing them to be very close to $\Lambda$CDM (see Figure \ref{fig:Chambeta} and \ref{fig:fR}). We also notice that the optical depth ($\tau$) and amplitude of scalar power spectrum ($A_s$) is higher for eChameleon gravity. This predicts higher redshift of reionization ($z_{re}=14.43 \pm 3.77$) and higher growth. In such a situation the only way in which the model can align itself with the measured $f\sigma_8$ is by choosing a smaller coupling parameter ($\beta_1$).
\noindent We have placed very stringent bounds on  $f(R)$ models with a   $\Lambda$CDM bacgkround, constraining their only free parameter to be $B_0 < 1.36 \times 10^{-5} $ (1 $\sigma$ C.L.). This is competitive with the constraint from solar system tests and clusters \citep{Hu2007,Schmidt2009} and other cosmological measurements \citep{XU2015, Dossett2014,Raveri2014, Planck2015DEMG}.  

Finally, we have analyzed the growth index  parametrization of the growth rate, measuring $\gamma =  0.612 \pm 0.072$ (11.7\% measurement), which is  completely consistent with the general relativity prediction. This is a slight improvement on  the 16\% measurement of \citet{Samushia14}. We also note that our measurement of growth index is slightly less precise than  current best measurement $\gamma =  0.665 \pm 0.0669$ (10\% measurement) \cite{Johnson2015} using combination of galaxy power spectrum, velocity power spectrum, Type {\rm I}a SNe, the cosmic microwave background (CMB), CMB lensing, and the temperature-galaxy cross correlation.

It is remarkable to notice that even after allowing many different kind of degree of freedom. our analysis shows that everything is consistent with vanilla $\Lambda$CDM cosmology and General theory of relativity.

\section*{Acknowledgement}
We would like to thank Martin White and Eric Linder for providing us many important suggestions on the draft of our paper and discussing various theoretical and observational details. We also want to thank Lado Samushia for feedback on the clarity of the text. We would also like to thank Mark Trodden and Levon Pogosian  for reading through the draft of our paper. 
S.A. and S.H. are supported by NASA grant NASA 12-EUCLID11-0004 and NSF AST1412966 for this work.
 A.S. acknowledges support from The Netherlands Organization for Scientific Research (NWO/OCW), and also from the D-ITP consortium, a program of the Netherlands Organisation for Scientific Research (NWO) that is funded by the Dutch Ministry of Education, Culture and Science (OCW).
This work made extensive use of the NASA Astrophysics Data System and of the {\tt astro-ph} preprint archive at {\tt arXiv.org}.